\documentclass{article}

\usepackage{proof}
\usepackage{latexsym}
\usepackage{url}
\usepackage{color}
\usepackage{graphics}

\newcounter{lab}

\newcommand{\nickname}{{\it Lyre}}

\newcommand{\bsc}[3]{\ensuremath{[#1;#2;#3]}}
\newcommand{\bscord}[4]{\ensuremath{[#1;#2;#3;#4]}}

\newcommand{\vbsc}[3]{\ensuremath{\langle #1;#2;#3 \rangle}}
\newcommand{\vbscord}[4]{\ensuremath{\langle #1;#2;#3;#4\rangle}}
\newcommand{\mix}[2]{\ensuremath{#1 \leftarrow #2}}
\newcommand{\red}[3]{\ensuremath{#1 \triangleleft #2 \triangleright #3}}

\newcommand{\freeze}[2]{\ensuremath{{\it freeze}_{#1}(#2)}}
\newcommand{\close}[1]{\ensuremath{{\it close}(#1)}}
\newcommand{\hide}[2]{\ensuremath{{\it hide}_{#1}(#2)}}

\newcommand{\rar}{\ensuremath{\rightarrow}}
\newcommand{\astr}{\ensuremath{\ast}}
\newcommand{\lar}{\ensuremath{\leftarrow}}

\newcommand{\mkmap}[1]{\ensuremath{\langle #1 \rangle}}
\newcommand{\Names}{{\it Names}}
\newcommand{\Ids}{{\it Idents}}
\newcommand{\IdsC}{{\it IdsC}}
\newcommand{\addmap}[2]{\ensuremath{#1 + #2}}
\newcommand{\maphide}[2]{\ensuremath{#1\backslash #2}}
\newcommand{\addord}[2]{\ensuremath{#1 \iaddord #2}}
\newcommand{\iaddord}{\ensuremath{\cup}}

\newcommand{\ext}[1]{\ensuremath{\overline{#1}}}
\newcommand{\extsf}[1]{\ensuremath{\overline{{\sf #1}}}}
\newcommand{\ixt}[1]{\ensuremath{\hat{#1}}}
\newcommand{\ixtsf}[1]{\ensuremath{\hat{{\sf #1}}}}

\newcommand{\ordmix}[4]{\ensuremath{\nu(#1, #2, #3, #4)}}
\newcommand{\ordclose}[2]{\ensuremath{\mu(#1, #2)}}
\newcommand{\iordmix}{\ensuremath{\nu}}
\newcommand{\iordclose}{\ensuremath{\mu}}

\newcommand{\exclose}[1]{\ensuremath{{\sf close}(#1)}}

\newcommand{\setrm}[2]{\ensuremath{#1 \backslash #2}}
\newcommand{\notlocked}[2]{\ensuremath{#1 \not\in\hspace{-.7ex}\in #2}}
\newcommand{\notconstrained}[2]{\ensuremath{#1 \not\prec #2}}

\newcommand{\evalmexpr}[4]{\ensuremath{#1 \vdash #2 \downarrow (#3; #4)}}
\newcommand{\evalexpr}[6]{\ensuremath{#1;#2 \vdash #3 \downarrow (#4; #5; #6)}}

\newcommand{\lc}{\ensuremath{l}}
\newcommand{\Hp}{\ensuremath{\sigma}}
\newcommand{\Loc}{{\it Loc}}
\newcommand{\ord}{\ensuremath{\theta}}
\newcommand{\Ord}{\ensuremath{\Theta}}
\newcommand{\trg}{\ensuremath{\pi}}
\newcommand{\Trg}{\ensuremath{\Pi}}
\newcommand{\sset}{\ensuremath{\delta}}
\newcommand{\SSet}{\ensuremath{\Delta}}

\newcommand{\seq}[3]{\ensuremath{#1 \stackrel{#2}{\mapsto} #3}}
\newcommand{\error}{{\tt error}}
\newcommand{\fresh}{{\rm fresh}}
\newcommand{\cmp}[2]{\ensuremath{#1\circ#2}}
\newcommand{\cmpd}[3]{\ensuremath{#1\circ#2\circ#3}}
\newcommand{\update}[3]{\ensuremath{#1[#2 \mapsto #3]}}
\newcommand{\updateseq}[4]{\ensuremath{#1[#2 \stackrel{#3}{\mapsto} #4]}}
\newcommand{\substseq}[4]{\ensuremath{#1[#4/#2]_{#3}}}

\newcommand{\dom}[1]{\ensuremath{{\it dom}(#1)}}
\newcommand{\cdom}[1]{\ensuremath{{\it cod}(#1)}}
\newcommand{\range}[1]{\ensuremath{{\it ran}(#1)}}

\newcommand{\emp}{\ensuremath{\emptyset}}
\newcommand{\FSharp}{F\#}
\newcommand{\mX}{\ensuremath{\mathcal{X}}}

\newcommand{\CMS}{{\it CMS}}

\title{Lazy mixin modules and disciplined effects}
\author{Keiko Nakata}

\begin{document}
\maketitle

\begin{abstract}
Programming languages are expected to support programmer's effort 
to structure program code. The ML module system, object systems 
and mixins are good examples of language constructs 
promoting modular programming. Among the three, 
mixins can be thought of as a generalization of the two others
in the sense that mixins can incorporate features of ML modules and objects
with a set of primitive operators with clean semantics.
Much work has been devoted to build mixin-based module systems 
for practical programming languages. 
In respect of the operational semantics, previous work notably investigated
mixin calculi in call-by-name and call-by-value evaluation settings.
In this paper we examine a mixin calculus in a call-by-need, or lazy, 
evaluation setting. 
We demonstrate how lazy mixins can be interesting in practice 
with a series of examples, and formalize the operational semantics 
by adapting Ancona and Zucca's concise formalization of call-by-name mixins.
We then extend the semantics with constraints to control 
the evaluation order of components of mixins in several ways. 
The main motivation for considering the constraints is to produce
side effects in a more explicit order than in a purely lazy, demand-driven
setting. We explore the design space of possibly interesting constraints
and consider two examples in detail. 
\end{abstract}

\section{Introduction}
Modularity is an important factor in the development of large programs. 
In particular programmer's effort to logically organize program code 
is of great importance in the long run to maintain, 
debug and extend the program code. 
Many modern programming languages have mechanisms to support this effort 
by facilitating modular development of programs. 
Examples of these mechanisms are object systems and the ML module system.

Being ML programmers, we enjoy the rich expressivity of 
the ML module system for modular programming. 
Nestable structures allow us to hierarchically organize namespaces 
and program code. With signature constraints, we control 
visibility of components of structures. In particular the combination of
nesting and signature constraints offers fine grained visibility
control, as witnessed, for example, by the Moby programming
language~\cite{classofmoby}. 
Functors, which are functions on modules, facilitate code
reuse in a modular way. 
Although functors might not be as pervasive as nesting or
signature constraints in our programs, 
they play a critical role in some contexts. 
Good examples are Map.Make and Set.Make functors 
as implemented in OCaml's standard library. 

However ML does not have recursive modules. That is, 
neither recursive functions nor types can be defined across 
module boundaries. As a result of this constraint, 
programmers may have to consolidate conceptually separate
components into a single module, intruding on modular
programming~\cite{recstruct}. 
Compared to ML, object systems have
excellent support for recursion across object/class 
boundaries.
Particularly in the presence of late-binding, object systems facilitate 
the development of extensible programs, where recursively defined types
and functions may need to be extended together.
Typical such scenarios are condensed in the notorious expression 
problem~\cite{expr_rev}. 

We expect a module system to support all the familiar features of ML modules
as well as recursion between modules and late-binding. 
There are at least three approaches to design such a module system.
Two of them are to extend the ML module system with recursion~\cite{Whatrecmod} 
and to extend object systems with nesting, abbreviation and type 
members
\cite{nuObj}. The third approach is to develop another form of
a module system, namely mixins. In this paper we follow this third approach.

\vspace{.5ex}

The concept of mixins is first introduced in the context of 
object systems \cite{jigsaw},
then is extended to in the context of ML-style modules. 
A mixin is a collection of named components, where each component can
be either defined (bound to a definition) or deferred (declared
without definition). Two key operations on mixins are the sum 
and freeze operations; the former 
takes two mixins and composes a new mixin by merging the two, and 
the latter resolves, or links, deferred components of a mixin to defined ones. 
The sum operation is reminiscent of functor application in ML 
and inheritance in object systems. However more flexibility is obtained
by separating the resolution from the sum operation. The freeze operation
is free to resolve a deferred component to a defined one 
independently of their names at any point, 
as long as their types match; notably
it can liberally tie a recursive knot inside a mixin. 
Indeed mixins are designed to be a generalization of ML-style modules
and objects, by incorporating features of both with a set of primitive
operators with clean semantics~\cite{jigsaw,cms}. 

Two challenging problems remain to replace ML-style modules with mixins,
namely type checking and initialization. 
In this paper, we address the latter problem. 
Promising progress has been made in designing a type system for
mixins with type components 
and a signature language to enforce type abstraction between 
mixins~\cite{nuObj,modules_et_units}; we expect to 
benefit from the previous work for the former problem. 

Initialization of mixins poses an important design problem 
if we are to build a mixin-based module system on top 
of a call-by-value core language supporting arbitrary side-effects, 
such as the ML core language. 
The main difficulty stems from unconstrained recursive
definitions such as {\sf let rec x = x} 
that the core language does not allow
but that might result from the freeze operation. 
We need an initialization semantics for mixins which 
takes account of unconstrained recursion consistently 
with the call-by-value semantics of the core language;
this is the subject of this paper. 

\vspace{.5ex}

Ancona and Zucca's mixin calculus~\cite{cms}, called \CMS, is one of the
most influential work in formalizing an operational semantics for
mixins. The formalization is elegantly concise and 
the paper illustrates how mixins
support various existing constructs for composing modules, found in the
ML module system, object-systems and linking calculi~\cite{linking}, 
well-explaining
why we claim mixins are a generalization of other modular systems. 
Inspired by \CMS, 
Hirschowitz and Leroy examined a mixin calculus in a call-by-value
setting to build a mixin-based module system on top of the ML core language.
The original call-by-name semantics of \CMS\
might not be well-suited for the call-by-value effectful ML core
language; \CMS\ admits recursive definitions such as 
{\tt let rec x = x}, causing the evaluation to diverge when
{\tt x} is selected, and it can produce the same side-effect
repeatedly. Neither behavior is consistent with the semantics of ML. 

In this paper we examine a mixin calculus in a call-by-need, or lazy, 
evaluation setting with a back-patching semantics.
Broadly we model mixins as nestable records 
with lazy fields. 
In the simplest setting which we examine
in Section~\ref{sec:lazymixins}, a component of a mixin is evaluated when it is
selected. We explain how side-effects are duplicated and linearly produced
by {\it open} and {\it closed} mixins respectively, while allowing both open
and closed mixins to be merged indiscriminately; 
this tolerance of merging mixins of different status
is our important design choice, which is different from previous 
proposals \cite{cbvmixins,cmsdo}.
In Section~\ref{sec:makeset}, we exemplify how the tolerance can be useful. 

Then in Section~\ref{sec:disciplined}, we extend the former semantics of 
Section~\ref{sec:lazymixins} 
with an ability to constrain the evaluation order of components of mixins. 
While the former semantics 
admits the most flexible recursive initialization patterns for mixins,
the flexibility can do harm because of the intrinsically
implicit evaluation order determined at run-time in a demand-driven way.
It may be interesting to constrain possible initialization patterns, making the 
evaluation order more explicit. For
instance, we may want to enforce top-down evaluation order within mixins,
where components of a mixin are evaluated following the textual definition order in
the source program. Or, we may want to keep the invariant that
once a component of a mixin is selected, all its components are eventually 
evaluated.  
Indeed our ultimate goal is to find the most beneficial constraint on the
evaluation order that still admits interesting recursive
initialization patterns, but that makes the evaluation order more explicit, 
thus more predictable to programmers. 

In this paper we do not propose a particular constraint; simply 
we do not yet have enough experience in programming with mixins to decide what
constraint is best in practice. Hence we formalize the operational semantics
so that it deals with several constraints. 
Concretely, we give a constraint language and 
extend the semantics of Section~\ref{sec:lazymixins} to evaluate components of mixins 
according to a given constraint expressed by the constraint language. 
We explore the design space of possibly interesting constraints 
and examine two particular constraints in detail as examples. 

\vspace{.5ex}

Contributions of the paper are summarized as follows.
We formalize the operational semantics for a lazy mixin calculus 
(Section~\ref{sec:lazymixins}), by moderately extending
Ancona and Zucca's formalization 
of a call-by-name mixin calculus~\cite{cms}.
We demonstrate how lazy mixins can be useful in practice through 
examples (Section~\ref{sec:examples}). 
Then we extend the semantics to be able to control 
the evaluation order of components of mixins in several
ways (Section~\ref{sec:disciplined}) and exemplify concrete
scenarios where particular evaluation strategies are enforced by constraints. 
We believe the ability to deal with several evaluation strategies is
a novelty of the formalization 
and the formalization serves as a basis for exploring 
the design space. 

\section{Examples}\label{sec:examples}

In this section, we introduce our lazy mixin calculus 
through a series of examples. Many constructs of the calculus come
from \CMS~\cite{cms}. 
Examples are written in a more programmer-friendly 
surface syntax and we assume a small subset of the OCaml core language
for the core language of the mixin calculus. 
We recall that the OCaml core language adopts a call-by-value evaluation strategy
and supports arbitrary side-effects within (core) expressions. 


\subsection{{\sf MakeSet} and {\sf MakeMultiSet} mixins}\label{sec:makeset}
\begin{figure}
{\sf \begin{tabular}{l}
module MakeSet = functor
(X~:~sig \\
~~val create : unit \rar\ int val compare : int \rar\ int \rar\ int end) \rar \\
~~struct\\
~~~~let create () = [ X.create () ]\\
~~~~let compare s1 s2 = ..... X.compare ...\\
~~end\\
module MakeMultiSet = functor(X~:~sig\\
~~val create : unit \rar\ int 
val compare : int \rar\ int \rar\ int end) \rar\\ 
~~struct\\
~~~~let create () =  [[ X.create () ]] \\
~~~~let compare s1 s2 = .... X.compare ... \\
~~end\\
module Key = struct \\
~~let count = ref (-1)\\
~~let create () = incr count; !count\\
~~let compare x y = \\
~~~~if x = y then 0 else if x $<$ y then 1 else (-1)\\
end\\
module Set = MakeSet(Key)\\
module MultiSet = MakeMultiSet(Key)\\
\end{tabular}}
\caption{{\sf MakeSet} and {\sf MakeMultiSet} ML functors}\label{functors_fig}
\end{figure}

We start by looking at possible mixin equivalents to 
{\sf MakeSet} and {\sf MakeMultiSet} ML functors 
and their instances by the {\sf Key} structure, 
as given in Figure~\ref{functors_fig}. 
{\sf MakeSet} and {\sf MakeMultiSet} are functions on ML structures, or functors, 
for making sets and multi-sets 
of integers, internally represented as lists and lists of lists respectively. 
Both take as argument a structure containing a {\sf create} function
for producing integers out of nothing and 
a {\sf compare} function for comparing given two integers.
{\sf MakeSet} extends this functionality to sets of integers and 
{\sf MakeMultiSet} does for multi-sets. 
Then we apply the two functors to the {\sf Key} structure to instantiate 
customized {\sf Set} and {\sf MultiSet} structures. 

Below is a mixin equivalent to the {\sf MakeSet} functor:\vspace{1ex}\\
\hspace*{5ex}{\sf \begin{tabular}{l}
mixin MakeSet = \{ \\
~~val create\_element : unit \rar\ int\\
~~val compare\_element : int \rar\ int \rar\ int\\
~~let create () = [ create\_element () ]\\
~~let compare s1 s2 = ..... compare\_element ... ~~~\}\\
\end{tabular}}\vspace{1ex}\\
A {\it mixin structure} is a sequence of named components, where a component may be 
defined like {\sf create} and {\sf compare} or deferred like 
{\sf create\_element} and {\sf compare\_element}.
The bodies of defined components can refer 
to names of deferred components as if they were present. 
Similarly, below is a mixin equivalent to the {\sf MakeMultiSet} functor:\vspace{1ex}\\
\hspace*{5ex}{\sf \begin{tabular}{l}
mixin MakeMultiSet = \{ \\
~~val create\_element : unit \rar\ int\\
~~val compare\_element : int \rar\ int \rar\ int\\
~~let create () = [[ create\_element () ]]\\
~~let compare s1 s2 = ..... compare\_element ... ~~~\}\\
\end{tabular}}\vspace{1ex}\\
We build a mixin equivalent to the {\sf Key} structure in two steps:\vspace{1ex}\\
\hspace*{5ex}{\sf \begin{tabular}{l}
mixin FKey = \{\\
~~let count = ref (-1)\\
~~let create\_key () = incr count; !count\\
~~let compare\_key x y = \\
~~~~if x = y then 0 else if x $<$ y then 1 else (-1) ~~\}\\
mixin Key = \exclose{{\sf FKey}}\\
\end{tabular}}\vspace{1ex}\\ 
We distinguish two states of mixins: open and closed. 
Intuitively a closed mixin is a record with lazy fields, 
whereas an open mixin is a function which returns a record with lazy fields. 
Intended implications of this comparison are 
1) projection of components is only possible from closed mixins, 
but not from open mixins;
2) an open mixin can be instantiated to create closed mixins; 
3) side-effects contained in a closed mixin are produced exactly once, 
whereas side-effects in an open mixin are produced as many times as
the mixin is instantiated. 
{\sf FKey} above is an open mixin, 
and {\sf Key} is a closed mixin instantiated from {\sf FKey} by the {\it close}
operation. A projection {\sf Key.create\_key} 
is legal, but {\sf FKey.create\_key} is not.
As expected, the counter {\sf Key.count} is initialized to
{\sf -1} exactly once. 

We may close the {\sf FKey} mixin again:\vspace{1ex}\\
\hspace*{3ex}{\sf mixin Key2 = close(FKey)}\vspace{1ex}\\
Counters {\sf Key.count} and {\sf Key2.count} are distinct from 
each other. For instance, calling {\sf Key.create} does not 
increase {\sf Key2.count}. 

The close operation is only available for mixins without holes,
or mixins which do not contain deferred components. 
The {\it freeze} operation, possibly combined with the {\it sum} operation, 
is used to fill in the holes. 
For instance, below we merge {\sf Key} and {\sf MakeSet} mixins by the sum operation, 
then resolve the deferred components {\sf create\_element} and 
{\sf compare\_element} of {\sf MakeSet} to the defined
component {\sf create\_key} and {\sf compare\_key} of {\sf Key} respectively by the
freeze operation, to fill in the holes of {\sf MakeSet}:\vspace{1ex}\\
{\sf \begin{tabular}{l}
mixin FSet' = \mix{{\sf Key}}{{\sf MakeSet}}\\
mixin FSet = freeze$_{\psi}$(FSet')
\end{tabular}}\vspace{1ex}\\
where $\psi$ is  the following mapping\footnote{Only in this section, 
we use angle brackets instead of square brackets to denote mappings 
to avoid confusion with list expressions.}:\vspace{1ex}\\
{\sf\begin{tabular}{l}
$\langle$ create\_element $\mapsto$ create\_key;\\
~ compare\_element $\mapsto$ compare\_key $\rangle$
\end{tabular}}\vspace{1ex}\\
Now {\sf FSet} does not contain holes, hence we can close it:\vspace{1ex}\\
\hspace*{1ex}{\sf mixin Set = \exclose{{\sf FSet}}}\vspace{1ex}\\
We do the same for {\sf MakeMultiSet} in one step this time:\vspace{1ex}\\ 
{\sf \begin{tabular}{l}
mixin MultiSet = close(freeze$_{\psi}$(Key \lar\ MakeMultiSet))
\end{tabular}}\vspace{1ex}\\
{\sf Set} and {\sf MultiSet} mixins are equivalent to 
{\sf Set} and {\sf MultiSet} ML structures 
as given in Figure~\ref{functors_fig}. 

It is important to see differences resulting from closing {\sf FKey} after
merging it with {\sf MakeSet} and {\sf MakeMultiSet} as follows:\vspace{1ex}\\
{\sf \begin{tabular}{l}
mixin Set' = close(freeze$_{\psi}$(FKey \lar\ MakeSet))\\
mixin MultiSet' = close(freeze$_{\psi}$(FKey \lar\ MakeMultiSet))\\
\end{tabular}}\vspace{1ex}\\
Here {\sf FKey} is instantiated twice. 
While {\sf Set} and {\sf MultiSet} share the same {\sf counter},
{\sf Set'} and {\sf MultiSet'} have distinct {\sf counter}'s. 
As a result, calling {\sf Set.create} affects 
both results of the next calls
to {\sf Set.create} and {\sf MultiSet.create}, while
calling {\sf Set'.create} only does the result of the next call to itself. 

\subsection{Widget mixins}\label{widget_mixins_sec}
Many GUI programming APIs involve recursive initialization patterns 
to form mutually referential graphs among related widgets
in the style of {\it create-and-configure}, 
as Syme calls in~\cite{initgraphs}. 
In that paper he explains how recursive initialization patterns 
are prominent in GUI programming
and proposes a semi-safe lazy evaluation strategy for the ML core language 
to tame ML's value recursion restriction; the restriction constrains 
right-hand sides of recursive definitions to be syntactic values, 
thus may hinder uses of sophisticated GUI programming APIs. 
In this subsection, we consider a GUI example 
similar to his in a lazy mixin setting.

We assume given the following interface of the API:\vspace{1ex}\\
{\sf\begin{tabular}{l}
type form type formMenu type menuItem \\
val createForm: string \rar\ form\\
val createMenu: string \rar\ formMenu\\
val createMenuItem: string \rar\ menuItem\\
val toggle : menuItem \rar\ unit\\
val setMenus : form \astr\ formMenu list \rar\ unit\\
val setMenuItems : formMenu \astr\ menuItem list \rar\ unit\\
val setAction : menuItem \astr\ (unit \rar\ unit) \rar\ unit
\end{tabular}}\vspace{1ex}\\
The API requires the create-and-configure initialization pattern where
widgets are first created, then explicit mutation configures a relation between 
the widgets.
Below we build  boiler-plate mixins which encapsulate 
the create-and-configure pattern:\vspace{1ex}\\
{\sf \begin{tabular}{l}
mixin Form = \{ \\
~~val name : string\\
~~val menus : formMenu list \\
~~let form = createForm(name)\\
~~let \_ = setMenus(form, menus) ~\}\\
 mixin Menu = \{ \\
~~val name : string\\
~~val items : menuItem list\\
~~let menu = createMenu(name)\\
~~let \_ = setMenuItems(menu, items) ~\}\\
mixin MenuItem = \{\\
~~val name : string\\
~~val other : menuItem\\
~~let item = createMenuItem(name)\\
~~let \_ = setAction(item, fun () \rar\ toggle(other)) ~\}
\end{tabular}}\vspace{1ex}\\
Then we use them:\vspace{1ex}\\
{\sf\begin{tabular}{l}
MyForm = \\
~~hide$_{{\sf name}}$(freeze$_{\mkmap{\sf name \mapsto name}}$(Form \lar\ \{ let name = ``Form'' \}))\\
MyMenu = \\
~~hide$_{{\sf name}}$(freeze$_{\mkmap{\sf name \mapsto name}}$(Menu \lar\ \{ let name = ``Menu'' \}))\\
MyItem1 = \\
~~rename$_{(\mkmap{\sf other \mapsto item2}, \mkmap{item1 \mapsto item})}$(hide$_{{\sf name}}$(\\
~~~~freeze$_{\mkmap{\sf name \mapsto name}}$(MenuItem \lar\ \{ let name = ``Rice'' \})))\\
MyItem2 = \\
~~rename$_{(\mkmap{\sf other \mapsto item1}, \mkmap{item2 \mapsto item})}$(hide$_{{\sf name}}$(\\
~~~~freeze$_{\mkmap{\sf name \mapsto name}}$(MenuItem \lar\ \{ let name = ``Grape'' \})))\\
MyGUI = \\
~~close(freeze$_{\psi_1}$(\\
~~~~MyItem1 \lar\ (MyItem2 \lar (MyMenu \lar MyForm))))\\
\end{tabular}}\vspace{1ex}\\  
where $\psi_1$ is the following mapping:\vspace{1ex}\\
\begin{tabular}{l}
$\langle$ {\sf item1} $\mapsto$ {\sf item1}; {\sf item2} $\mapsto$ {\sf item2}; \\
~ {\sf items} $\mapsto$ {[}{\sf item1}; {\sf item2}{]} ; {\sf menus} $\mapsto$ {[}{\sf menu}{]}$\rangle$
\end{tabular}
\vspace{1ex}\\
Above we have introduced two new constructs.
The {\it hide} operation ${\sf hide}_{X}(M) $
hides the component named $X$ of the mixin $M$ by making the component 
invisible outside.
The {\it rename} operation ${\sf rename}_{(\phi_1, \phi_2)}(M)$
changes names of deferred and defined components of the mixin $M$ by
$\phi_1$ and $\phi_2$ respectively, where 
$\phi_1$ and $\phi_2$ are finite mappings on names of 
components\footnote{In the formalization, we will distinguish
$\alpha$-convertible {\it identifiers} (internal names)  
and non-convertible {\it names} (external names). 
Then $\phi_1$ and $\phi_2$ are mappings on names, not on identifiers.}. 
For instance, in the definition of {\sf MyItem1}, the deferred component {\sf other}
is renamed to {\sf item2} and the defined component {\sf item} to {\sf item1}.
Observe the opposite directions of mappings \mkmap{{\sf other \mapsto item2}} for 
deferred components
and \mkmap{{\sf item1 \mapsto item}} for defined ones. 
This adds flexibility of the rename operation in that
when $\phi_1$ maps deferred components of distinct names to the same name,
then the components can be resolved simultaneously, and that when 
$\phi_2$ maps two distinct names $X_1$ and $X_2$ to a single name $X$, 
then the component which was named $X$ can now projected by either $X_1$ or
$X_2$. 
Thanks to the renaming, {\sf MyItem1}
and {\sf MyItem2} can be merged to form {\sf MyGUI} without causing name clash;
a deferred component and a defined component of a mixin can have the same name,
but a deferred (resp. defined) component must not have the same name 
as other deferred (resp. defined) components. 
Besides, we have used the freeze operation in a more flexible way by
mapping a name of a deferred component
to an expression composed of names of defined components, 
for example [{\sf item1}; {\sf item2}]. 

The above example builds a GUI application forming a widget containment hierarchy
where {\sf MyForm} contains {\sf MyMenu}, which contains 
{\sf MyItem1} and {\sf MyItem2}. 
{\sf MyItem1} and {\sf MyItem2} are mutually recursive; 
each toggles the activation state of the other. 
We hide the {\sf name} component of mixins to 
be merged to avoid name clash. Anonymous (under-scored) components
do not contribute to name clash; they
would be implemented as syntax sugar via the hide operation.
Renaming of {\sf other} and {\sf item} components of {\sf MyItem1} and 
{\sf MyItem2}, both of which are derived from the same mixin {\sf MenuItem}, 
is necessary for cross-connecting 
the deferred component of one to the defined component of the other. 
In the last line we sum up the constituent mixins and 
resolve deferred components to defined ones, configuring the 
widget containment hierarchy. 

This example suggests it can be useful 
to control the evaluation order of components of mixins, 
instead of evaluating them purely lazily in a demand-driven way,
i.e. evaluating only the components that are projected.
Indeed we would like to make sure 
the widget containment hierarchy has been properly configured 
before {\sf MyGUI} is actively used. 
For that purpose, 
all the anonymous components of {\sf MyGUI} must be evaluated
before any of its components becomes externally accessible. 
In Section~\ref{sec:disciplined},
we present the operational semantics which can enforce 
such constraint on the evaluation order.

\subsection{A combinator library for marshallers}\label{sec:picklers} 
The last example deals with marshaller combinators
and is motivated by Syme's paper again~\cite{initgraphs}.
Kennedy introduced a functional-language combinator library
for building marshallers and unmarshallers
of data structures \cite{pickler}. 
The essential ingredient of his proposal is the tying together
of a marshaller and unmarshaller pair in a single value. 
Then the consistency of marshalling and unmarshalling
is ensured by construction.  
The original proposal of Kennedy is implemented in Haskell.
Porting the code to ML is mostly easy, except for a couple of wrinkles. 
The value recursion restriction of ML 
is a source for the wrinkles and requires cumbersome workaround
for an ML version of the combinator. 
In this subsection, we rebuild a combinator library for marshallers
using lazy mixins with ML as the core language and demonstrate how
the use of lazy mixins avoids the value recursion problem. 
In the following examples we assume a richer mixin language where a mixin 
may contain deferred and defined types as components, although 
our formal development does not consider mixins with type components. 
Type checking of mixins is not in the scope of the paper.
As far as our examples are concerned, 
existing type systems are sufficient~\cite{cool&hot}. 

We consider a mixin-based combinator library with 
the specification given below. 
A mixin signature 
{\sf mixin M : \{ type t ~val x : t \} \rar\ \{ type s = t \astr\ t ~val y : s \}}
specifies an open mixin with deferred type component 
{\sf t} and value component {\sf x} of type {\sf t}, 
written in the left-hand side of the 
arrow, and with defined type component {\sf s} satisfying type equation
{\sf s = t \astr\ t} and value component 
{\sf y} of type {\sf s}, written in the right-hand side\footnote{This signature
language is designed only for the sake of the examples. A more practical
signature language is proposed, for instance in~\cite{modules_et_units}.}.
The scope of type names declared in the left hand of the arrow extends 
to the right hand. 
\vspace{1ex}\\
{\sf\begin{tabular}{l}
type channel\\
type $\alpha$ marshaller \\
val marshal : $\alpha$ marshaller \rar\ $\alpha$ \astr\ channel \rar\ unit\\
val unmarshal : $\alpha$ marshaller \rar\ channel \rar\ $\alpha$\\
mixin PairMrshl : \\
~\{ type s1 type s2 \\ 
~\hspace*{2ex}val mrshl1 : s1 marshaller ~val mrshl2 : s2 marshaller \} \rar \\
~\{ type t = s1 \astr\ s2 ~val marshaller : t mrshl  \} \\
mixin ListMrshl : \\
~\{ type elm ~val mrshl\_elm : elm marshaller \} \rar \\
~\{ type t = elm list ~val mrshl : t marshaller  \} \\
mixin InnerMrshl : \\
\{ type src ~type trg ~val f : src \rar\ trg ~val g : trg \rar\ src\\
\hspace*{1.5ex}  val mrshl\_src : src marshaller \}
\rar \\
\{ type t = trg ~val mrshl : t marshaller \}\\
mixin IntMrsh : \\
~\{\} \rar\ \{ type t = int ~val mrshl : t marshaller \} \\
mixin StringMrsh :\\
~\{\} \rar\ \{ type t = string ~val mrshl : t marshaller \}
\end{tabular}}\vspace{1ex}\\ 
{\sf IntMrsh} and {\sf StringMrsh} mixin have an empty deferred component.
But they are still open mixins, thus needs to be closed for making 
their components accessible. 

The abstract type {\sf marshaller} could be internally implemented 
as a record consisting of a marshalling action
and unmarshalling action:\vspace{.5ex}\\
{\sf \begin{tabular}{rl}
type $\alpha$ marshaller = 
&\{ marshal: $\alpha$ \astr\ channel \rar\ unit; \\
&\hspace*{1.5ex} unmarshal : channel \rar\ $\alpha$ \}
\end{tabular}}\vspace{.5ex}\\
Recall that it is important for consistently building 
marshallers and unmarshallers that 
a {\sf marshaller} is a single value, but not two separate functions.
In this way, users of the library can only build consistent 
marshaller/unmarshaller pairs. 

We do not present further details of how the library can be implemented.
In the original paper~\cite{pickler}, 
Kennedy explains an excellent implementation which lets
the programmer control sharing of the marshaled data. 
Transposing marshaller combinators for constructed types such as 
pairs and lists, originally implemented as functions, to
mixins is straightforward. For instance, a function-based 
marshaller combinator {\sf pairMrshl(mrshl1, mrshl2)} of type
{\sf (s1 marshaller) \astr\ (s2 marshaller) \rar\ (s1 \astr\ s2) marshaller}
for constructing a marshaller for a pair from marshallers
of the components can be translated into a mixin as follows:
\vspace{.5ex}\\
{\sf\begin{tabular}{l}
mixin PairMrshl = \{\\
~type s1 type s2 \\
~val mrshl1 : s1 marshaller val mrshl2 : s2 marshaller \\
~type t = s1 \astr\ s2 \\
~let mrshl = (\astr\ the body of pairMrshl(mrshl1, mrshl2) \astr) \}\\
\end{tabular}}\vspace{1ex}

Now we turn to how to build custom-marshallers for user-defined 
data types. 
We first build marshallers for both a single file, represented as a pair of 
an integer and string, and a list of files:\vspace{.5ex}\\
{\sf \begin{tabular}{l}
type file = int \astr\ string\\
mixin FileMrshl = freeze*( \\
~~(rename$_{(\emp,\mkmap{{\sf s1\mapsto t; mrshl1 \mapsto mrshl}})}$(IntMrshl)) \lar\\
~~(rename$_{(\emp, \mkmap{{\sf s2\mapsto t; mrshl2 \mapsto mrshl}})}$(StringMrshl)) \lar\ 
{\sf PairMrshl}))\\
mixin FilesMrshl = freeze*(\\
~~(rename$_{(\emp, \mkmap{{\sf elm \mapsto t; mrshl\_elm \mapsto mrshl}})}$(FileMrshl)) \lar\
ListMrshl)
\end{tabular}}\vspace{.5ex}\\ 
The notation \emp\ denotes an empty mapping. 
Above we have introduced a high-level mixin construct {\sf freeze*},
which resolves deferred components to the same-named defined components 
if exists, then hides the defined components used. 
The formalization given in the next section does not include {\sf freeze*}.
For the surface language, we could implement it by combining 
freeze and hide operations with the help of the type system. 

Next we build marshallers for both a single folder and a list of folders,
which form recursive data structures:\vspace{.5ex}\\
\hspace*{3ex}{\sf \begin{tabular}{l}
type folder = \{ files: file list; subfldrs: folders \}\\
and folders = folder list\\
\end{tabular}}\vspace{.5ex}\\
Like Syme, we use an intermediate mixin in favor of conciseness. 
\vspace{.5ex}\\ 
{\sf \begin{tabular}{l}
mixin FldrInnerMrshl = freeze*(\\
~~\{ type src = folder type trg = folders\\
~~\hspace*{1.5ex}  let f (fls, fldrs) = \{ files = fls; subfldrs = fldrs \}\\
~~\hspace*{1.5ex}  let g fld = (fld.files, fld.subfldrs) \} \lar\\
~~InnerMrshl) \\
\end{tabular}}\vspace{.5ex}\\
Then to build marshallers for {\sf folder} and {\sf folders},
we follow their type definitions,
by merging constituent mixins and resolving deferred components to defined ones:
\vspace{.5ex}\\
{\sf \begin{tabular}{l}
mixin FldrMrshl = close(freeze*(\\
~~(rename$_{(\emp,\mkmap{{\sf s1\mapsto t; mrshl1 \mapsto mrshl}})}$(FilesMrshl)) \lar\\
~~(hide$_{{\sf t}}$(rename$_{(\mkmap{{\sf s1 \mapsto s1;s2 \mapsto s2; mrshl1 \mapsto mrshl1;mrshl2 \mapsto mrshl2}}}$, \\
\hspace{15ex}{}$_{\mkmap{{\sf mrshl\_src \mapsto mrshl}})}$(PairMrshl))) \lar\\
~~(rename$_{({\mkmap{{\sf mrshl\_src \mapsto mrshl\_src}}},\mkmap{{\sf elm\mapsto t; mrshl\_elm \mapsto mrshl; fldMrshl \mapsto mrshl}})}$\\
\hspace{10ex}(FldrInnerMrshl)) \lar\\
~~(rename$_{(\mkmap{{\sf elm \mapsto elm; mrshl\_elm \mapsto mrshl\_elm}},}$\\
\hspace{10ex}{}$_{\mkmap{{\sf fldrsMrshl \mapsto mrshl; mrshl2 \mapsto mrshl;  s2\mapsto t}})}$(ListMrshl))
\end{tabular}}\vspace{.5ex}\\
Thus we have created marshaller {\sf FldrMrshl.fldrMrshl} for a single folder
and {\sf FldrMrshl.fldrsMrshl}  for a list of folders. 
In the surface language we could provide an identity mapping with
appropriate domain to improve notational verbosity. 
For the above example, we preferred not to use identity mappings to 
make explicit how deferred components are resolved to defined ones.

\vspace{1ex}

Specifying marshallers for recursive data types such as {\sf folder} and
{\sf folders} is not problematic in the original Haskell context of Kennedy 
or in our lazy mixin context, essentially due to the laziness.
But it is problematic in the context of the ML core language because of
the value recursion restriction.
The problem is well-discussed in the previous papers of Kennedy and Syme.
In short, it is due to ML's intolerance of the following recursive 
definition\footnote{The example is not ideal in that 
ML does not allow recursive type definitions such as {\sf type t = t \astr\ int}. 
But the source of the problem should be clear.}
\vspace{.5ex}\\
\hspace*{3ex}{\sf let rec p = pairMrshl(p, intMrshl)}
\vspace{.5ex}\\
where we assumed primitive marshaller {\sf intMrshl} and 
marshaller combinator {\sf pairMrshl} given in the library. 
Note that the abstraction of type {\sf marshaller} is not the root of the problem.
Even we exposed the underlying implementation of {\sf marshaller},
we still face the problem since marshallers are {\it pairs} of functions. 

\vspace{1ex}

If we consider first-class mixins, i.e. mixins as core values, 
we could implement a marshaller as separate functions of 
marshal action and unmarshal action, while ensuring the consistency of 
constructed marshaller/unmarshaller pairs.
Our formalization of a lazy mixin calculus is largely abstracted over
the core language. 
The core language may support first-class mixins, 
but we preferred the current presentation in favor of generality by
not assuming a richer core language. We also found the current presentation
useful to demonstrate a possibly interesting scenario where 
lazy mixins and a call-by-value core language are combined 
to obtain more tolerant recursive definitions.

\section{Lazy mixins}\label{sec:lazymixins}

We start by considering a simpler semantics, 
where a component of a mixin is evaluated when it is projected.
The syntax of our lazy mixin calculus, named \nickname, is defined in
Figure~\ref{syntax_lazy_fig}. 
We assume pairwise disjoint sets 
\Ids\ of identifiers, \Names\ of names, and \Loc\ of locations. 
Components of a mixin are internally referred to by 
($\alpha$-convertible) identifiers, but externally accessed by 
(non-convertible) names. We use locations to formalize lazy evaluation.

\begin{figure}
\begin{tabular}{lcll}
$x, y$ &$\in$& \Ids & {\it identifiers}\\
$X, Y$ &$\in$& \Names & {\it names}\\
$l$ &$\in$& \Loc & {\it locations}\vspace{1ex}\\
$E$ &::=& $C \mid M$ &{\it expressions}\\
$M$ &::=& \bsc{\iota}{o}{\rho} & {\it mixin structure}\\
 &$\mid$& \mix{M_1}{M_2} & {\it sum}\\
 &$\mid$& \red{\phi_1}{M}{\phi_2} & {\it rename}\\
 &$\mid$& \hide{X}{M} & {\it hide} \\
 &$\mid$& \freeze{\psi}{M} & {\it freeze}\\
 &$\mid$& \close{M} & {\it close}\\
 &$\mid$& $M.X$ & {\it projection}\\
 &$\mid$& $x$ & \\
 &$\mid$& $X$ & \\
 &$\mid$& $\lc$ & {\it location}\\
$C$ &::=&   $M.X \mid \lc $ & {\it projection, location} \\
    &$\mid$& $x \mid X $ & {\it identifier, name}\\
    &$\mid$& $\ldots$\\
$\iota$ &::=& \seq{x_i}{i \in I}{X_i} & {\it input assignment}\\
$o$ &::=& \seq{X_i}{i \in I}{x_i} & {\it output assignment}\\
$\rho$ &::=& \seq{x_i}{i \in I}{E_i} & {\it local binding} \\
$\phi$ &::=& \seq{X_i}{i \in I}{Y_i} & {\it renaming}\\
$\psi$ &::=& \seq{X_i}{i \in I}{E_i} & {\it tying}
\end{tabular}
\caption{Syntax for \nickname}\label{syntax_lazy_fig}
\end{figure}

\vspace{1ex}

{\it Notations}
For a finite mapping $f$, \dom{f} and \range{f} 
respectively denote the domain and range of $f$. 
\emp\ is an empty mapping, that is, \dom{\emp},  \cdom{\emp} and  
\range{\emp} are empty sets. 
The notation \seq{a_i}{i\in I}{b_i} denotes the finite mapping $f$ such 
that, for all $i\in I$, $f(a_i) = b_i$. It is only defined when,
for all $i, j \in I$, $i \not= j$ implies $a_i \not= a_j$. 
Throughout the paper, we only consider finite mappings, 
so simply say a mapping to mean a finite one. 

For a mapping $f$, \maphide{f}{x} is the restriction of $f$ to 
$\dom{f} \backslash \{ x \}$. 


For mappings $f, f'$, we write \addmap{f}{f'} for 
the union of $f$ and $f'$. That is, 
\dom{\addmap{f}{f'}} = $\dom{f} \cup \dom{f'}$,
and for all $x$ in \dom{\addmap{f}{f'}},\\
$(\addmap{f}{f'})(x) = \left\{ \begin{array}{ll}
                   f(x) &\mbox{when} ~x \in \dom{f}\\
                   f'(x) &\mbox{when} ~x \in \dom{f'}
                     \end{array}\right.$\\
\addmap{f}{f'} is defined only if $\dom{f} \cap \dom{f'} = \emptyset$. 
The notation \cmp{f'}{f} denotes the mapping composition. 
It is defined only when $\range{f} \subseteq \dom{f'}$.

\vspace{1ex}

We have explained most of the constructs for mixin expressions
in the previous section. 
In the formalization, however, mixin structures take a more fundamental form. 
Precisely, a mixin structure, simply called structure hereafter, 
is a triple of {\it input assignment} $\iota$, 
{\it output assignment} $o$ and {\it local binding} $\rho$.
The local binding $\rho$ is a mapping from identifiers to expressions
and corresponds to the body of the mixin; 
if $x$ is in \dom{\rho},
then $x$ is a defined component of the mixin 
with $\rho(x)$ being the defining expression. 
$\rho(x)$ can refer to both defined and deferred components 
of the mixin via identifiers. Any identifier in 
\dom{\rho} and \dom{\iota} is bound in $\rho(x)$. 
$\rho(x)$ must not contain names.
The input assignment $\iota$ is a mapping from identifiers to names
and corresponds to declarations of deferred components; 
a deferred component internally referred to by $x$ is resolved by the name 
$\iota(x)$. 
\dom{\rho} and \dom{\iota} must be disjoint. 
The output assignment $o$ is a mapping from names to identifiers; 
a component of a mixin externally accessed by $X$ is associated to $o(X)$
inside the mixin. 
\range{o} must be a subset of $\dom{\rho} \cup  \dom{\iota}$. 
Structures are identified up to $\alpha$-renaming of identifiers.
The explicit distinction between identifiers and names allows 
identifiers to be renamed by $\alpha$-conversion, 
while names remain immutable, thus making projection 
by name unambiguous~\cite{cbvmixins,HOMwithsharing}. 

In agreement with the distinction of identifiers and names,
the rename operation takes two mappings on names,
and the hide operation takes a name as argument.
The freeze operation takes a mapping from names to expressions, 
where expressions may contain names. 
Mixin expressions contain locations. 
We use locations in the operational semantics to implement
lazy evaluation, but locations will not appear in the surface language.

The formalization is mostly independent of the core language. 
We only assume the core language includes
projection from mixin expressions, locations, identifiers
and names. Again locations will not appear in the surface language. 

\vspace{1ex}

\begin{figure}
\begin{tabular}{c}
\begin{tabular}{lcll}
$V$ &::=& $\vbsc{\iota}{o}{\rho} \mid v$ & {\it values}\\
$\kappa$ &::=& $E \mid \error\ \mid\ V$  &   {\it heap objects}\\  
$\Hp$ &$\in$& $\Loc$ ~-fin-$>$ ~$\kappa$ & {\it heap state} \\
\end{tabular}
\vspace{2ex}\\
\begin{tabular}{c}
{\it mixin structure}\\
\refstepcounter{lab}
\infer[(\thelab\label{eval_bsc})]{
  \evalmexpr{\Hp}{\bsc{\iota}{o}{\rho}}{\vbsc{\iota}{o}{\rho}}{\Hp}
}{}\vspace{.5ex}\\
{\it sum}\\
\refstepcounter{lab}
\infer[(\thelab\label{eval_sum})]{
  \evalmexpr{\Hp}{\mix{M_1}{M_2}}
            {\vbsc{\addmap{\iota_1}{\iota_2}}{\addmap{o_1}{o_2}}
                  {\addmap{\rho_1}{\rho_2}}}{\Hp_3}
}{
  \deduce{
    \dom{\rho_1} \cap \dom{\rho_2} = \emp
    ~~~\dom{\iota_1} \cap \dom{\iota_2} = \emp
  }{
    \evalmexpr{\Hp}{M_1}{\vbsc{\iota_1}{o_1}{\rho_1}}{\Hp_2}
    &\evalmexpr{\Hp_2}{M_2}{\vbsc{\iota_2}{o_2}{\rho_2}}{\Hp_3}
  }
}
\vspace{.5ex}\\
{\it rename}
\vspace{.5ex}\\
\refstepcounter{lab}
\infer[(\thelab\label{eval_rename})]{
  \evalmexpr{\Hp}{\red{\phi}{M}{\phi'}}
            {\vbsc{\cmp{\phi}{\iota}}
                  {\cmp{o}{\phi'}}{\rho}}
            {\Hp_2}
}{
  \deduce{
  }{
    \evalmexpr{\Hp}{M}{\vbsc{\iota}{o}{\rho}}{\Hp_2}
  }
}
\vspace{.5ex}\\
{\it hide}
\vspace{.5ex}\\
\refstepcounter{lab}
\infer[(\thelab\label{eval_hide})]{
  \evalmexpr{\Hp}{\hide{X}{M}}{\vbsc{\iota}{\maphide{o}{X}}{\rho}}{\Hp_2}
}{
  \evalmexpr{\Hp}{M}{\vbsc{\iota}{o}{\rho}}{\Hp_2}
}
\vspace{.5ex}\\
{\it freeze}
\vspace{.5ex}\\
\refstepcounter{lab}
\infer[(\thelab\label{eval_freeze})]{
  \evalmexpr{\Hp}{\freeze{\psi}{M}}
            {\vbsc{\iota_2}{o}{(\addmap{\rho}{(\cmpd{o}{\psi}{\iota_1})}}}
            {\Hp_2}
}{
  \deduce{
    \dom{\psi} = \range{\iota_1}
    ~~\dom{\psi} \cap \range{\iota_2} = \emp
  }{
    \evalmexpr{\Hp}{M}{\vbsc{\addmap{\iota_1}{\iota_2}}{o}{\rho}}{\Hp_2}
  }
}\vspace{.5ex}\\
{\it close}
\vspace{.5ex}\\
\refstepcounter{lab}
\infer[(\thelab\label{eval_close})]{
  \evalmexpr{\Hp}{\close{M}}{\vbsc{\emp}{o}{\seq{x_i}{i\in I}{\lc_i}}}
            {\updateseq{\Hp_2}{\lc_i}{i\in I}
                       {\substseq{E_i}{x_j}{j\in I}{\lc_j}}}
}{
  \evalmexpr{\Hp}{M}{\vbsc{\emp}{o}{\seq{x_i}{i\in I}{E_i}}}{\Hp_2}
  &\forall i \in I,\ \lc_i ~\fresh
}
\vspace{.5ex}\\
{\it projection}
\vspace{.5ex}\\
\refstepcounter{lab}
\infer[(\thelab\label{eval_proj})]{
  \evalmexpr{\Hp}{M.X}{V}{\Hp_3}
}{
  \evalmexpr{\Hp}{M}{\vbsc{\iota}{o}{\rho}}{\Hp_2}
  &\evalmexpr{\Hp_2}{(\cmp{\rho}{o})(X)}{V}{\Hp_3}
}
\vspace{.5ex}\\
{\it location}
\vspace{.5ex}\\
\refstepcounter{lab}
\infer[(\thelab\label{eval_loc_val})]{   
  \evalmexpr{\Hp}{\lc}{V}{\Hp}
}{
  \Hp(\lc) = V
}
\vspace{1ex}\\
\refstepcounter{lab}
\infer[(\thelab\label{eval_loc_expr})]{   
  \evalmexpr{\Hp}{\lc}{V}{\update{\Hp_2}{\lc}{V}}
}{
 \Hp(\lc) = E
 &\evalmexpr{\update{\Hp}{\lc}{\error}}{E}{V}{\Hp_2}
}
\end{tabular}
\end{tabular}
\caption{Semantics}\label{eval_lazy_fig}
\end{figure}

In Figure~\ref{eval_lazy_fig}, 
we define the operational semantics for \nickname. 
A {\it heap state} \Hp\ is a mapping from locations to {\it heap objects},
which are either expressions, values or \error. 
We formalize lazy evaluation by suspending and memorizing evaluation 
in heap states. We assume given core values $v$.
Then values $V$ are either structures or core values. 
We syntactically distinguish structures as mixin expressions, 
surrounded by square brackets, and as values, 
surrounded by angle brackets. 
The distinction lets us simplify the formalization. 

The judgment \evalmexpr{\Hp}{E}{V}{\Hp_2} means that 
in heap state \Hp\ expression $E$ evaluates into value $V$ with 
heap state being $\Hp_2$. 
We assume given inference rules to 
deduce \evalmexpr{\Hp}{C}{V}{\Hp_2} for core expressions other than
projections or locations. 

{\it Notations} 
We write \updateseq{\Hp}{\lc_i}{i\in I}{\kappa_i} to 
denote a mapping extension. Precisely,
\vspace{1ex}\\
\hspace*{5ex}\updateseq{\Hp}{\lc_i}{\i \in I}{\kappa_i}($\lc'$) = $\left\{ \begin{array}{ll}
\kappa_i & ~~{\rm when} ~\lc' = \lc_i\ \mbox{for some} ~i \in I\\
\Hp(\lc') & ~~{\rm otherwise}\\
\end{array}\right. $\vspace{1ex}\\
The notation is defined only if,
for any $i, j \in I$, $i \not= j$ implies $\lc_i \not= \lc_j$. 
We may write \update{\Hp}{\lc}{\kappa} when $I$ is a singleton. 
The notation \substseq{E}{x_i}{i\in I}{\lc_i} denotes 
the substitution of $l_i$'s for $x_i$'s in $E$
for all $i \in I$. The notation is defined only if,
for any $i, j \in I$, $i \not= j$ implies $x_i \not= x_j$. 
For input assignment $\iota$, tying $\psi$ and output assignment 
$o = \seq{X_i}{i\in I}{x_i}$,
\cmpd{o}{\psi}{\iota} is 
the mapping $\rho$ such that \dom{\rho} = \dom{\iota}
and, for all $x \in \dom{\iota}$, 
$\rho(x)$ is the expression obtained from $\cmp{\psi}{\iota}(x)$ by replacing
$X_i$'s with $x_i$'s for all $i \in I$. 
\cmpd{o}{\psi}{\iota} is only defined when $\range{\iota} \subseteq \dom{\psi}$
and, for all $x \in \dom{\iota}$, 
any name appearing in $\cmp{\psi}{\iota}(x)$ is in \dom{o}. 

\vspace{1ex}

Let's look at the inference rules. 
A structure evaluates into itself and the heap state is unchanged 
(rule (\ref{eval_bsc})).
The sum operation merges the two operand mixins (rule (\ref{eval_sum})). 
The side conditions ensure that identifiers do not collide 
and are always satisfiable by taking appropriate $\alpha$-equivalent mixins. 
The effect of the rename operation is simply compositions 
of mappings (rule (\ref{eval_rename})). 
The hide operation narrows the domain of the output assignment
(rule (\ref{eval_hide})). 
The freeze operation resolves deferred components according 
to the tying $\psi$ (rule (\ref{eval_freeze})).
Precisely, for all $x \in \dom{\iota_1}$, $x$ is resolved to the expression
$\cmpd{o}{\psi}{\iota_1}(x)$. $\cmp{\psi}{\iota_1}(x)$ 
may contain names, which are replaced  with identifiers by $o$. 
The rule augments the local binding with 
\cmpd{o}{\psi}{\iota_1} and 
the input assignment of the resulting mixin diminishes accordingly. 

The rule (\ref{eval_close}) for the close operation  is 
responsible for making mixins lazy. 
Firstly the operand mixin expression
must be evaluated into a structure without holes.
Then, for each defined identifier $x_i$, a fresh heap location $\lc_i$
is allocated to store the defining expression $E_i$, 
where any occurrence of $x_j$'s in $E_i$ is substituted by $\lc_j$'s.
The body of the resulting structure maps $x_i$ to 
$\lc_i$, thus access to $x_i$ is redirected to $\lc_i$.

To evaluate projection $M.X$ (rule (\ref{eval_proj})), 
$M$ is first evaluated into 
a structure \vbsc{\iota}{o}{\rho}. 
Then the rule consults $o$ 
for the associated identifier to $X$, 
thus determines the expression that $M.X$ accesses by looking up the identifier
in $\rho$. In the normal evaluation, i.e. when projection is made from
a closed mixin, $\cmp{\rho}{o}(X)$  returns a location, 
where the defining expression of $o(X)$ is stored. 

The last two rules are for evaluating locations \lc,
and are fairly standard. 
If the heap state contains a value at \lc\ (rule (\ref{eval_loc_val})), 
then the value is returned and the heap state is unchanged.
When an expression $E$ is stored at \lc, then it is evaluated. 
The heap state maps \lc\ to \error\ during the evaluation of  $E$, 
to avoid evaluating the same expression repeatedly 
and to signal an error for cyclic definitions. 
On completion, the heap state is updated with the resulting value. 

It may be useful to note that we do not need evaluation rules 
for names or identifiers, since in the normal evaluation 
names are substituted by expressions and identifiers by locations. 

As one may have noticed, 
there is potential that evaluation gets stuck.
One serious source is when projection is made from an open mixin. 
Instead of preventing such a scenario at the operational semantics level,
we leave it to the type system to eliminate the possibilities for evaluation 
to get stuck. 
Although we do not present a type system in this paper,
straightforward adaptation of previous work such as \cite{cool&hot} 
is sufficient for this purpose. The type system would be able to 
eliminate other ill-typed scenarios such as an attempt to merge 
mixins with overlapping output names, i.e., 
$\dom{o_1} \cap \dom{o_2} \not= \emptyset$ in rule (\ref{eval_sum}). 
We have omitted inference rules for propagating error states,
assuming when evaluation encounters an \error\ during
deduction, it immediately terminates signaling a runtime error.

\subsection{Call-by-name and eager variants}\label{variants_sec}

Small modifications to evaluation rules let the operational semantics 
model call-by-name and eager evaluation strategies. 
This subsection presents those variants. 

To model a call-by-name strategy, we replace rules 
(\ref{eval_loc_val}) and (\ref{eval_loc_expr}) with 
the single rule:\vspace{1ex}\\
\hspace*{\fill}
\refstepcounter{lab}
\infer[(\thelab\label{evalcbn_close})]{
  \evalmexpr{\Hp}{\lc}{V}{\Hp_2}
}{
  \evalmexpr{\Hp}{\Hp(\lc)}{V}{\Hp_2}
}\hspace*{\fill}\vspace{1ex}\\
Heap states are not updated, thus expressions stored 
are re-evaluated whenever accessed.

By eager evaluation strategy, we mean an evaluation strategy that evaluates all components
of a mixin at once when the mixin is closed. This is also easily implemented
by replacing rule (\ref{eval_close}) with:\vspace{1ex}\\
\refstepcounter{lab}
\[
\infer[(\thelab\label{evalcbv_close})]{
  \evalmexpr{\Hp}{\close{M}}{\bsc{\emp}{o}{\seq{x_i}{i\in I}{\lc_i}}}
            {\Hp'_n}
}{
  \deduce{
    \evalmexpr{\Hp'}{\lc_1}{V_1}{\Hp'_1}
    ~~~\evalmexpr{\Hp_1'}{\lc_2}{V_2}{\Hp'_2}
    ~~\cdots
    ~~\evalmexpr{\Hp'_{n-1}}{\lc_n}{V_n}{\Hp'_n}
  }{
    \deduce{
      \forall i \in I,\ \lc_i ~\fresh
      ~~\Hp' = \updateseq{\Hp_2}{\lc_i}{i\in I}{\substseq{E_i}{x_j}{j\in I}{\lc_j}}
    }{
      \evalmexpr{\Hp}{M}{\vbsc{\emp}{o}{\seq{x_i}{i\in I}{E_i}}}{\Hp_2}
      &I = \{1, 2, \ldots, n\}
    }
  }
}\]

\section{Lazy mixins and disciplined effects}\label{sec:disciplined}

We extend the operational semantics of the previous section so that
it takes account of constraints on the evaluation order of
components of mixins. 
In examples of this section, 
we use a side-effecting function ``{\sf print} $C$'', which
prints the resulting value of evaluating $C$, 
then returns the value, to visualize the evaluation order. 

\subsection{Design space of evaluation strategies}\label{sec:strategies}
There are several evaluation strategies that we found interesting to consider. 
Below we explain those strategies which we have in mind.

Firstly we consider the following example:\vspace{.5ex}\\
{\sf \begin{tabular}{l}
mixin M1 = close(\\
~~\{ let c1 = print 1 ~let c2 = print 2\\
\hspace*{.5ex}~~~~let c3 = print 3 ~let c4 = print 4 \})\\
\end{tabular}}\vspace{.5ex}\\
{\it Top-down strategy}
We may want components of a mixin to be evaluated following 
the textual definition order in which they appear in the source program. 
This constraint will ensure, in the above example, that
1 is printed before 2, and 2 is before 3, and 3 is before 4. 
This strategy is reminiscent of ML's strategy. \\
{\it Lazy-field and lazy-record strategies}
If we adopt top-down strategy, we have a choice
on whether to make accessible 
components of a mixin immediately after they are evaluated, 
while other components of the mixin are still to be evaluated consecutively. 
We call lazy-field strategy the strategy that allows such access
and lazy-record strategy the strategy that does not allow.
Intuitively lazy-field strategy treats a closed mixin as a record with lazy fields like  
{\sf \{ a1 = lazy (print 1); a2 = lazy (print 2) \}}, where ``\{`` and ``\}''
are record constructors of the core language here, 
while lazy-record strategy does as 
a lazy record like {\sf lazy (\{ a1 = print 1; a2 = print 2 \})}. 
For instance, let's consider executing the program: \vspace{1ex}\\
{\sf \begin{tabular}{l}
mixin M2 = close (\{ let c1 = 1 ~let c2 = 2 \astr\ M3.c1 \})\\
mixin M3 = close (\{ let c1 = 3 + M2.c1 \})\\
let main = M2.c2\\
\end{tabular}}\vspace{1ex}\\
We assume, for the explanatory purpose, that 
a program consists of a sequence of mixin definitions plus a core value 
definition named {\sf main}, where the top-level mixin bindings 
can be mutually recursive. 
The execution of the program evaluates 
the defining expression of {\sf main}. 
For the execution of the above program to succeed, 
{\sf M2.c1} should be accessible to {\sf M3.c1} before 
evaluation of {\sf M2.c2} is completed. 
Hence the execution succeeds with lazy-field strategy, 
but it fails with lazy-record strategy. 
Clearly lazy-field strategy is more permissive. 
Although restrictive, lazy-record strategy is interesting to consider 
particularly in the presence of finer grained accessibility control, 
as we will see below. \\
{\it Internal and external accessibilities}
It can be useful to change accessibility to components of a mixin 
depending on whether the access is made inside the same mixin or outside. 
For instance, let us consider the following program:\vspace{1ex}\\
{\sf \begin{tabular}{l}
mixin M4 = close(\\ 
~~\{ let c1 = 1 + 2  ~let c2 = c1 + 4 ~let c3 = print ``ok'' \})\\
let main = M4.c2
\end{tabular}}\vspace{1ex}\\
For the execution to succeed, {\sf M4.c1} should be accessible to 
{\sf M4.c2} immediately after  {\sf M4.c1} is evaluated but before 
evaluation of {\sf M4.c2} is completed. 
Hence we need lazy-field strategy inside {\sf M4}. 
However we may want components of {\sf M4} to become accessible outside, 
only after all components of {\sf M4} have been evaluated. 
In other words, we may want lazy-record strategy outside {\sf M4} 
to make sure ``{\sf ok}'' is necessarily printed 
before any component of {\sf M4} becomes externally accessible. 
Indeed, in the widget mixins example of Section~\ref{widget_mixins_sec},
we envisaged this internally-lazy-field externally-lazy-record strategy.
For instance, the component {\sf menu} of {\sf MyGUI} should be 
accessible inside once it is created, to configure the 
widget containment hierarchy via functions {\sf setMenuItems}
and {\sf setMenus}. However {\sf menu} should not be accessed outside {\sf MyGUI},
before the configuration is completed, i.e. all components, including 
anonymous ones, of {\sf MyGUI} are evaluated. 
Internally-lazy-field externally-lazy-record strategy is also 
potentially interesting in the light of our experience in programming 
with ML modules; it is sometimes
convenient to include anonymous side-effecting expressions 
at the end of a structure for the debugging purpose or to properly initialize 
mutable components of the structure. 

\vspace{1ex}

As well as design choices among above strategies, 
we have a choice on how to propagate constraints when merging or
closing mixins. 
In contrast to this vast design space,
we do not have enough experience in programming with
mixins. Plainly we cannot determine at present which 
strategy is most beneficial in practice. 
Hence 
we keep our formalization open to strategies. That is,
the formalization is abstracted over constraints on strategies and 
can be instantiate to express particular strategies. 

\subsection{The constraint language}

We use binary relations as our constraint language 
to express various evaluation strategies. 
We identify binary relations on any set $P$ as 
sets of pairs of elements in $P$. 
For instance, the relation
{\sf \{(c1, c2) (c2, c3) (c3, c4)\}} expresses 
top-down strategy in the first example. 
The relation {\sf \{(c1, c3) (c2, c4)\}} stipulates that {\sf c1} should be 
evaluated before {\sf c3} and 
{\sf c2} before {\sf c4}. There is no constraint on the evaluation 
order between {\sf c1} and {\sf c2} or {\sf c3} and {\sf c2}.
Hence this relation does not fix the evaluation order in a unique way.
For instance in the first example, either of the output ``1 2 3 4'' or `` 2 1 3 4''
is compatible with the relation. 

To deal with more expressive strategies,
we distinguish three sorts of identifiers: 
(ordinary) identifiers $x$ for controlling the evaluation order;
{\it internal identifiers} \ixt{x} for internal accessibility; 
{\it external identifiers} \ext{x} for external accessibility. 
For instance, the relation $\{(x_1, \ixt{x_2})\}$ 
stipulates that the component $x_1$ must be 
evaluated before the component $x_2$ becomes accessible inside the same mixin.
Similarly $\{(x_1, \ext{x_2})\}$ stipulates 
that $x_1$ must be evaluated before $x_2$ becomes accessible outside. 
We will formalize how relations on these three sorts of identifiers 
control the evaluation order and accessibility later, when we define
the operational semantics. 
Below we give a descriptive explanation 
on how those three sorts of identifiers can be used 
to express strategies we proposed above.

If we assign the relation {\sf \{(c1, c2)\}} as the constraint 
to {\sf M2} 
in the second example, the execution of the program succeeds. 
The relation stipulates nothing about how 
{\sf M2.c1} becomes accessible, 
implying {\sf M2.c1} is accessible both inside {\sf M2} and outside
immediately after it is evaluated. 
If we assign the relation
{\sf \{(c1, c2), (c2, \extsf{c1})\}} as the constraint to {\sf M2}, 
the execution fails. The relation stipulates that
{\sf M2.c1} becomes accessible outside {\sf M2} only 
after {\sf M2.c2} has been evaluated. However {\sf M3.c1} needs to 
access {\sf M2.c1} to be evaluated, and {\sf M2.c2} to {\sf M3.c1};
there is a circular dependency. 

Top-down internally-lazy-field externally-lazy-record strategy in the third example
is expressed by assigning the relation
{\sf \{(c1, c2), (c2, c3), (c3, \extsf{c1}), (c3, \extsf{c2})\}} as
the constraint to {\sf M4}.
We remark that there are implicit constraints imposed by the semantics,
such as {\sf (c1, \ixtsf{c1})} and {\sf (c1, \extsf{c1})} 
(but not {\sf (\ixtsf{c1}, \extsf{c1})}). 
The reason is simple: 
the component {\sf c1} must be evaluated before it becomes accessible.
Hence if the relation included {\sf (\extsf{c1}, c1)}, 
which introduces a circular dependency together with the implicit constraint
{\sf (c1, \extsf{c1})}, the evaluation would fail. Precisely,
the operational semantics signals an error. 
The relation 
{\sf \{(c1, c2), (c2, c3), (c3, \ixtsf{c1}), (c3, \ixtsf{c2})
(c3, \extsf{c1}), (c3, \extsf{c2})\}} in the third example
expresses top-down internally-lazy-record externally-lazy-record strategy. 
We have omitted including constraints such as {\sf (c2, \ixtsf{c1})},
since it is anyway induced from {\sf (c2, c3)} and  {\sf (c3, \ixtsf{c1})}
by transitivity. Transitivity is imposed by the semantics: 
if {\sf c2} must be evaluated before {\sf c3}, and {\sf c3} before
\ixtsf{c1}, then naturally {\sf c2} must be evaluated before \ixtsf{c1}.

\begin{figure*}
\scalebox{.8}{\begin{tabular}{c}
{\it mixin structure}\\
\refstepcounter{lab}
\infer[(\thelab\label{evalcnstr_bsc})]{
  \evalexpr{\Hp}{\Ord}{\bscord{\iota}{o}{\rho}{\ord}}
           {\vbscord{\iota}{o}{\rho}{\ord}}{\Hp}{\Ord}
}{}\vspace{.5ex}\\
{\it sum}\\
\refstepcounter{lab}
\infer[(\thelab\label{evalcnstr_sum})]{   
  \evalexpr{\Hp}{\Ord}{\mix{M_1}{M_2}}
           {\vbscord{\addmap{\iota_1}{\iota_2}}{\addmap{o_1}{o_2}}
                    {\addmap{\rho_1}{\rho_2}}
                    {\ordmix{\dom{\iota_1}\cup\dom{\rho_1}}{\ord_1}
                           {\dom{\iota_2}\cup\dom{\rho_2}}{\ord_2}}}
           {\Hp_3}{\Ord_3}
}{
  \deduce{
    \dom{\rho_1} \cap \dom{\rho_2} = \emp
    ~~~\dom{\iota_1} \cap \dom{\iota_2} = \emp
  }{
    \evalexpr{\Hp}{\Ord}{M_1}{\vbscord{\iota_1}{o_1}{\rho_1}{\ord_1}}{\Hp_2}{\Ord_2}
    &\evalexpr{\Hp_2}{\Ord_2}{M_2}{\vbscord{\iota_2}{o_2}{\rho_2}{\ord_2}}
              {\Hp_3}{\Ord_3}
  }
}
\vspace{.5ex}\\
{\it rename}
\vspace{.5ex}\\
\refstepcounter{lab}
\infer[(\thelab\label{evalcnstr_rename})]{   
  \evalexpr{\Hp}{\Ord}{\red{\phi_1}{M}{\phi_2}}
            {\vbscord{\cmp{\phi_1}{\iota}}{\cmp{o}{\phi_2}}{\rho}{\ord}}
            {\Hp_2}{\Ord_2}
}{
  \deduce{
  }{
    \evalexpr{\Hp}{\Ord}{M}{\vbscord{\iota}{o}{\rho}{\ord}}{\Hp_2}{\Ord_2}
  }
}
\vspace{.5ex}\\
{\it hide}
\vspace{.5ex}\\
\refstepcounter{lab}
\infer[(\thelab\label{evalcnstr_hide})]{
  \evalexpr{\Hp}{\Ord}{\hide{X}{M}}{\vbsc{\iota}{\maphide{o}{X}}{\rho}}{\Hp_2}{\Ord_2}
}{
  \evalexpr{\Hp}{\Ord}{M}{\vbsc{\iota}{o}{\rho}}{\Hp_2}{\Ord_2}
}
\vspace{.5ex}\\
{\it freeze}
\vspace{.5ex}\\
\refstepcounter{lab}
\infer[(\thelab\label{evalcnstr_freeze})]{   
  \evalexpr{\Hp}{\Ord}{\freeze{\psi}{M}}
           {\vbscord{\iota_2}{o}{\addmap{\rho}{(\cmpd{o}{\psi}{\iota_1})}}{\ord}}
           {\Hp_2}{\Ord_2}
}{
  \evalexpr{\Hp}{\Ord}{M}{\vbscord{\addmap{\iota_1}{\iota_2}}{o}{\rho}{\ord}}
           {\Hp_2}{\Ord_2}
  &\dom{\psi} = \range{\iota_1}
  &\dom{\psi} \cap \range{\iota_2} = \emp
}\vspace{.5ex}\\
{\it close}
\vspace{.5ex}\\
\refstepcounter{lab}
\infer[(\thelab\label{evalcnstr_close})]{   
  \evalexpr{\Hp}{\Ord}{\close{M}}
           {\vbscord{\emp}{o}{\seq{x_i}{i\in I}{\lc''_i}}
                    {\ordclose{\{x_i \mid i \in I\}}{\ord}}}
           {\updateseq{
              \updateseq{
                \updateseq{\Hp_2}{\lc_i}{i\in I}
                          {\substseq{E_i}{x_j}{j\in I}{\lc'_j}}}
                        {\lc'_i}{i\in I}{\lc_i}}
                      {\lc''_i}{i\in I}{\lc_i}}
           {\addord{\Ord_2}{\Ord_3}}
}{
  \deduce{   
    \forall i \in I, ~\lc_i ~\lc'_i ~\lc''_i ~\fresh
    ~~\Ord_3 = \substseq{\substseq{\substseq{\ord}{x_i}{i\in I}{\lc_i}}
                                  {\ixt{x_i}}{i\in I}{\lc'_i}}
                        {\ext{x_i}}{i\in I}{\lc''_i}
  }{
    \evalexpr{\Hp}{\Ord}{M}{\vbscord{\emp}{o}{\seq{x_i}{i\in I}{E_i}}{\ord}}
             {\Hp_2}{\Ord_2}
  }
}
\vspace{.5ex}\\
{\it projection}
\vspace{.5ex}\\
\refstepcounter{lab}
\infer[(\thelab\label{evalcnstr_proj})]{  
  \evalexpr{\Hp}{\Ord}{M.X}{V}{\Hp_3}{\Ord_3}
}{
  \evalexpr{\Hp}{\Ord}{M}{\vbscord{\iota}{o}{\rho}{\ord}}{\Hp_2}{\Ord_2}
  &\evalexpr{\Hp_2}{\Ord_2}{(\cmp{\rho}{o})(X)}{V}{\Hp_3}{\Ord_3}
}
\vspace{.5ex}\\
{\it location}
\vspace{.5ex}\\
\refstepcounter{lab}
\infer[(\thelab\label{evalcnstr_loc_val})]{   
  \evalexpr{\Hp}{\Ord}{\lc}{V}{\Hp}{\Ord}
}{
  \Hp(\lc) = V
}
\vspace{1ex}\\
\refstepcounter{lab}
\infer[(\thelab\label{evalcnstr_loc_expr_later})]{   
  \evalexpr{\Hp}{\Ord}{\lc}{V}{\Hp_3}{\Ord_3}
}{
    (\lc', \lc) \in \Ord
    &\Hp(\lc) = E
    &\evalexpr{\update{\Hp}{\lc}{\error}}
              {\setrm{\Ord}{\{(\lc', \lc)\}}}{\lc'}{V'}{\Hp_2}{\Ord_2}
    &\evalexpr{\update{\Hp_2}{\lc}{E}}{\Ord_2}{\lc}{V}{\Hp_3}{\Ord_3}
}
\vspace{1ex}\\
\refstepcounter{lab}
\infer[(\thelab\label{evalcnstr_loc_expr_now})]{   
  \evalexpr{\Hp}{\Ord}{\lc}{V}{\update{\Hp'}{\lc}{V}}{\Ord'}
}{
  \notconstrained{\lc}{\Ord}
  &\Hp(\lc) = E
  &\evalexpr{\update{\Hp}{\lc}{\error}}{\Ord}{E}{V}{\Hp'}{\Ord'}
}
\end{tabular}}
\caption{Semantics with constraint}\label{eval_disciplined_fig}
\end{figure*}

\subsection{Operational semantics}

We extend a structure with 
a binary relation on ordinary, internal and external identifiers. 
Precisely, a structure \bscord{\iota}{o}{\rho}{\ord} 
is now 4-tuple of an input assignment $\iota$, output assignment $o$,
local binding $\rho$, and {\it local constraint} \ord, where 
\ord\ is a binary relation on 
$ \dom{\iota} \cup \dom{\rho} \cup 
\{ \ext{x} \mid x \in \dom{\iota} \cup \dom{\rho}\}
\cup \{ \ixt{x} \mid x \in \dom{\iota} \cup \dom{\rho}\}$. 
Otherwise the syntax is unchanged from Figure~\ref{syntax_lazy_fig}. 
We use \mX\ as a metavariable
for sets of ordinary, internal and external identifiers.

We assume given two functions 
\ordclose{\mX_1}{\ord_1} and 
\ordmix{\mX_1}{\ord_1}{\mX_2}{\ord_2}, 
where $\ord_1$ and $\ord_2$ 
are binary relations on $\mX_1$ and $\mX_2$ respectively. 
\ordclose{\mX_1}{\ord_1} returns a binary relation on $\mX_1$ and
\ordmix{\mX_1}{\ord_1}{\mX_2}{\ord_2} does on $\mX_1 \cup \mX_2$. 
The operational semantics uses \ordclose{\mX_1}{\ord_1} to build 
local constraints for mixins instantiated by the close operation, 
where $\mX_1$ is the set of deferred and defined identifiers 
and $\ord_1$  the local constraint of the operand mixin. 
Similarly it uses \ordmix{\mX_1}{\ord_1}{\mX_2}{\ord_2} to build
local constraints for mixins composed by the sum operation, 
where $\mX_1$ (resp. $\mX_2$) is 
the set of deferred and defined identifiers 
and $\ord_1$ (resp. $\ord_2$) is the local constraints
of the left (resp. right) right operand mixin. 
By not fixing the interpretations of \iordclose\ or \iordmix, 
we keep the formalization neutral of how to propagate local constraints when 
closing or merging mixins. 

\vspace{1ex}

In Figure~\ref{eval_disciplined_fig}, we present the operational semantics
which takes account of constraints. 
We use \Ord\ as a metavariable for {\it global constraints}, 
or binary relations on locations. 
The judgment \evalexpr{\Hp}{\Ord}{E}{V}{\Hp'}{\Ord'} means that
under global constraint \Ord\ with heap state \Hp\ 
expression $E$ evaluates into $V$, where the heap state and 
global constraint have evolved into $\Hp'$ and $\Ord'$. 
As well as heap states, global constraints evolve during evaluation, 
since new constraints are added when a mixin is closed 
(rule (\ref{evalcnstr_close})).
The notation \setrm{\Ord}{\{(\lc', \lc)\}} denotes 
set subtraction. 
The notation \notconstrained{\lc}{\Ord} denotes the condition that
there is not $\lc'$ such that $(\lc', \lc) \in \Ord$. 

Compared to previous evaluation rules in Figure~\ref{eval_lazy_fig},
modifications are made on rules 
(\ref{eval_sum}) (\ref{eval_close}), and (\ref{eval_loc_expr}), 
which we explain in turn. 

As explained above,
the sum operation uses the function \iordmix\ to build 
a local constraint for the composed mixin (rule (\ref{evalcnstr_sum})).

When closing a mixin (rule (\ref{evalcnstr_close})), 
three fresh locations $\lc_i, \lc_i', \lc''_i$ 
are created for each defined component $x_i$ of the mixin, 
in order to separately control the evaluation order by $\lc_i$'s,
internal accessibility by $\lc'_i$'s, and 
external accessibility by $\lc''_i$'s.
Four important points to be understood are as follows: 
1) The local binding of the resulting mixin maps $x_i$'s to
$\lc''_i$'s, i.e. locations for external access; 
2) Evaluation of $E_i$ is suspended at  $\lc_i$, 
where $x_i$'s in $E_i$ are substituted by 
$\lc'_i$'s, i.e. locations for internal access; 
3) The new heap state maps $\lc'_i$'s and $\lc''_i$'s for 
internal and external accesses to $\lc_i$'s 
to connect the accesses to underlying expressions; 
4) The local constraint \ord\ is transported to the global constraint,
by instantiating ordinary, internal and external identifiers to 
the corresponding locations. The notation 
\substseq{\substseq{\substseq{\ord}{x_i}{i\in I}{\lc_i}}
                                  {\ixt{x_i}}{i\in I}{\lc'_i}}
                        {\ext{x_i}}{i\in I}{\lc''_i} denotes 
the binary relation on locations obtained from \ord\ by substituting
$\lc_i$'s for $x_i$'s , $\lc'_i$'s for \ixt{x_i}'s, and 
$\lc''_i$'s for \ext{x_i}'s, where the substitution is performed by regarding
a binary relation as a set of pairs. 
The local constraint for the resulting mixin 
is build by the function \iordclose. 

We have introduced two new rules 
(\ref{evalcnstr_loc_expr_later})
and (\ref{evalcnstr_loc_expr_now}) for evaluating locations, 
replacing the previous rule (\ref{eval_loc_expr}), 
to take the global constraint into account.
Rule (\ref{evalcnstr_loc_expr_later}) considers the case
where the global constraint stipulates that
$\lc'$ should be evaluated before $\lc$,
as described by the side-condition $(\lc', \lc) \in \Ord$.
The rule evaluates $\lc'$ first, while updating the heap state
to map $\lc$ to \error\ and removing $(\lc', \lc)$ from \Ord;
if the evaluation of $E'$ attempts 
to evaluate \lc\ against the global constraint, an error is signaled. 
On completion of the evaluation, $E$ is restored at \lc\ and 
the rule retries to evaluate \lc. 
The same rule (\ref{evalcnstr_loc_expr_later}) may be
applied again, when there is another location 
which should be evaluated before \lc. 
Otherwise rule (\ref{evalcnstr_loc_expr_now}) is applicable.
When there is no location which must be evaluated before $\lc$, 
as described by the side condition \notconstrained{\lc}{\Ord} in 
rule (\ref{evalcnstr_loc_expr_now}), then $E$ is evaluated immediately.

It shall be informative to note that
in rule (\ref{evalcnstr_loc_expr_later}) the heap state \Hp\ 
is updated with \error\ to enforce the evaluation order 
stipulated by the constraint, while in rule (\ref{evalcnstr_loc_expr_now}) 
to detect ill-founded recursion. An important ingredient of the formalization 
is that evaluation of projection $M.X$ always goes through a location. 
This facilitates to control the evaluation order of components of mixins,
in terms of a binary relation on locations. 
It is easily proved that if $\Hp(\lc) = V$ then \notconstrained{\lc}{\Ord}.
We also remark that the operational semantics, 
in particular the evaluation order, 
is not necessarily deterministic, depending on the constraint. 
For instance, if we assign a local constraint 
{\sf \{(c1, c3), (c2, c4)\}} to the mixin {\sf M1} of the first example 
in Section~\ref{sec:strategies}, we can build deductions which result in 
the outputs ``1 2 3 4'' and ``2 1 3 4''.
Non-determinism is not abnormal, but should be thought of as 
underspecification like the underspecified evaluation order 
of function parameters in some programming languages. 

The operational semantics can be proved sane in the sense that 
evaluation does not diverge due to ill-founded recursion. In other
words, when evaluation diverges, heap states are extended infinitely. 
A heap state is extended only when a mixin is 
closed (rule (\ref{evalcnstr_close})).
Hence the heap explosion implies that the close operation is used
infinitely often. This is similar to a situation 
where infinite recursion of function calls exhausts the stack. 

%

\subsection{An example}\label{sec:example_frozenmodules}
We present an example by instantiating 
the constraint to express one particular evaluation strategy. 
The strategy 
is motivated by our previous work on examining a lazy evaluation
strategy for recursive ML-style module~\cite{lazymodules}. 
The purpose of the example is not to advocate 
this particular strategy, but to deliver the better intuition 
about how to use the constraint. 

For a structure \bscord{\iota}{o}{\rho}{\ord}, 
we let \ord\ be the relation: \vspace{1ex}\\
$\begin{array}{l}
\{(x_i, x_j) \mid x_i \in \dom{\rho} \cap \IdsC, x_j \in \dom{\rho}, i < j \} ~\cup\\
\{(x_i, \ext{x_j}) \mid \\
~~x_i \in (\dom{\iota} \cup \dom{\rho}) \cap \IdsC, x_j \in \dom{\iota} \cup \dom{\rho}\}
\end{array}$\vspace{1ex}\\
Above we have assumed for any $x_i, x_j$, if $i < j$ then 
the definition or declaration for $x_i$ 
textually precedes that for $x_j$ in the source program. 
\IdsC\ denotes the set of identifiers bounds to core expressions.
I.e. we assume \Ids\ consists of disjoint sets of identifiers
to be bound to core expressions and mixin expressions, respectively. 

We give interpretations to functions \iordclose\ and \iordmix\ as
follows:\vspace{1ex}\\
\hspace*{1ex}$\begin{array}{l}
\ordclose{\mX}{\ord} = \emp\\
\ordmix{\mX}{\ord}{\mX'}{\ord'} = \\
~~\ord ~\cup\ \ord' ~\cup 
\{(x_i, \ext{x_j} \mid x_i \in (\mX \cup \mX') \cap \IdsC, x_j \in \mX \cup \mX' \} 
\end{array}$\vspace{1ex}\\

The strategy is close to top-down internally-lazy-field 
externally-lazy-record strategy, where sub-mixins are evaluated lazily. 
Below we explain the strategy in detail. 
\begin{enumerate}\itemsep=0ex
\item 
Preceding core components of the same and enclosing mixins are evaluated
first. Enclosing mixins may contain other sub-mixins, 
whose core fields need not be evaluated first.
This is enforced by including in the local constraint of a structure the relation 
$\{(x_i, x_j) \mid x_i \in \dom{\rho} \cap \IdsC, x_j \in \dom{\rho}, i < j\}$.
Note that $x_j$ may be bound to a mixin expression, and in order to evaluate
the expression all core components preceding to it must have been evaluated. 
Thus the constraint controls the evaluation order 
not only of core components of the same mixin
but also of those of enclosing mixins. 

\item 
Components of a mixin become accessible inside the same mixin 
immediately after they are evaluated, but are accessible outside only after 
all the core components have been evaluated. 
This is enforced by not having constraints mentioning internal identifiers
and by including in the local constraint of a structure the relation 
$\{(x_i, \ext{x_j}) \mid x_i \in (\dom{\iota} \cup \dom{\rho}) \cap \IdsC, x_j \in \dom{\iota} \cup \dom{\rho}\}$. 
Observe how we propagate the constraint on external accessibility 
when merging mixins by including the relation
$\{(x_i, \ext{x_j} \mid x_i \in (\mX \cup \mX') \cap \IdsC, x_j \in \mX \cup \mX' \} $ in the result of \iordmix. 

\item 
While core components are evaluated following the textual definition order, 
there is no constraint on the evaluation order between components of merged mixins; 
\iordmix\ does not introduce $(x_i, x_j)$ or $(x_i, x_j)$
for any $x_i \in \mX$ and $x_j \in \mX'$. 

\item We do not leave any constraint after a mixin is closed; 
it is enough to enforce the evaluate order once. 
This is reflected in the interpretation of \iordclose.

\end{enumerate}

As we have said, we do not intend to justify ourselves in choosing this strategy,
because we do not have enough programming experience to do so.
Nevertheless we motivate the strategy below, aiming at posing questions to readers 
about possible concerns one may face in the quest of better design choices. 

The first condition ensures that backward references to core components of
the same and enclosing mixins are necessarily proper values, but not 
suspensions or \error's. This provides programmers
with a safety guarantee on backward references to core components,
which seems useful for ML-initiated programmers. Interestingly 
this design choice also bears a similarity to Java's class initialization
policy~\cite{javaspec}. 
We have already motivated in Section~\ref{sec:strategies} 
the combination of internally-lazy-field and externally-lazy-record strategies;
internally-lazy-field strategy leaves flexibility 
in intra-mixin recursion, while externally-lazy-record strategy keeps
the evaluation stable towards outside. 
A possible drawback of externally-lazy-record strategy, compared to
externally-lazy-field strategy, is that we may loose 
flexibility in inter-mixin recursion. 
However we are less concerned by inter-mixin recursion, 
since we believe mixins are designated to support flexible 
intra-mixin recursion;
notably the sum and freeze operations are useful for taking fix-points inside a mixin.

We are less confident in the choice of \iordmix. 
But it could be useful to keep the evaluation order 
independent between separately defined mixins; it is unlikely that
a programmer can foresee in which order components of other mixins to be merged with 
should/could be evaluated. 

\subsection{An extension}\label{sec:extension}

There is a strategy which we want to consider, but our constraint language is
not expressive enough for it. The strategy is a variant on top-down 
internally-lazy-field externally-lazy-field strategy, where we impose an extra
constraint, called {\it trigger-constraint}, 
that all components of a mixin must be evaluated at once before 
the first access to a component of the mixin returns.
For instance, let us consider the following program:\vspace{1ex}\\
{\sf\begin{tabular}{l}
mixin M1 = \{ \\
~~let c1 = print 1 ~let c2 = M2.c2\\
~~let c3 = print (c1 + c2) ~let c4 = print 5 \}\\
mixin M2 = \{ \\
~~let c1 = M1.c1 ~let c2 = print (c1 + 1) ~let c3 = print 4 \}\\
let main = M1.c3 
\end{tabular}}\vspace{1ex}\\
According to the above proposed strategy, 
the execution succeeds and ``1 2 4 3 5'' is printed. 
(Recall that ``{\sf print (1+2)}'' 
returns 3, after printing 3). Here is why.
\begin{enumerate}\itemsep=0ex
\item Top-down strategy ensures the printing orders ``1 3 5'' and ``2 4'' independently.
\item Thanks to internally-lazy-field strategy, 
{\sf M1.c3} and {\sf M2.c2} are successfully evaluated. 
\item Thanks to externally-lazy-field strategy, evaluation of {\sf M2.c1} succeeds.
Note that {\sf M2} is forced to evaluate by the access 
from {\sf M1.c2}. Hence with externally-lazy-record strategy,
the evaluation fails, since {\sf M1.c1} is not 
yet accessible outside then. 
\item Finally and importantly, the trigger-constraint ensures that
{\sf M2.c3} is evaluated before evaluation of {\sf M1.c2} returns and
that {\sf M1.c4} is before evaluation of {\sf main} returns. 
This explains why 4 is printed immediately after 2, and 5 is after 3.
\end{enumerate}

This strategy comes from our previous work on lazy recursive 
modules~\cite{lazymodules}. 
In a recursive modules setting,
inter-module recursion is important, hence we wanted externally-lazy-field
strategy. We found the trigger-constraint useful to 
enforce our design policy that once a module is accessed, 
all its component are eventually evaluated. 
We want to consider the same strategy as our previous proposal 
in a mixin context, too.
Observe that the trigger-constraint only lets the access
to {\sf M1.c3} trigger evaluation of {\sf M1.c4}, but
{\sf M1.c1} is already accessible both inside and outside 
once it is evaluated. 
We cannot include any of constraints 
{\sf (c4, c3)},  {\sf (c4, \extsf{c3})}, or
{\sf (c4, \ixtsf{c3})} in the local constraint of {\sf M1},
since the first one is inconsistent with top-down strategy,
the second with externally-lazy-field strategy, 
the third with internally-lazy-field strategy.

\subsubsection{Formalization}
\begin{figure*}
\scalebox{.8}{\begin{tabular}{c}
\refstepcounter{lab}
\infer[(\thelab\label{evalmore_loc_expr_trigger})]{   
  \evalexpr{\Hp}{\Trg}{\lc}{V}{\Hp'_n}{\Trg'_n}
}{
  \deduce{
    ~~~~\evalexpr{\Hp'}{\Trg'}{\lc_1}{V_1}{\Hp'_1}{\Trg'_1}
    ~~~\evalexpr{\Hp_1'}{\Trg_1'}{\lc_2}{V_2}{\Hp'_2}{\Trg'_2}
    ~~\cdots
    ~~\evalexpr{\Hp'_{n-1}}{\Trg'_{n-1}}{\lc_n}{V_n}{\Hp'_n}{\Trg'_n}~~~
  }{
    \Trg = (\Ord, \SSet)
    & \{ \lc, \lc_1 \ldots \lc_n \} \in \SSet
    &\evalexpr{\Hp}{(\Ord,\setrm{\SSet}{\{\{ \lc, \lc_1 \ldots \lc_n \}\}})}{\lc}{V}{\Hp'}{\Trg'}
  }
}
\vspace{1ex}\\
\refstepcounter{lab}
\infer[(\thelab\label{evalmore_loc_expr_later})]{   
  \evalexpr{\Hp}{(\Ord, \SSet)}{\lc}{V}{\Hp_3}{\Trg_2}
}{
  \deduce{
    \evalexpr{\update{\Hp}{\lc}{\error}}
             {(\setrm{\Ord}{\{(\lc', \lc)\}}, \SSet)}{\lc'}{V'}{\Hp_2}{\Trg_2}
    ~~\evalexpr{\update{\Hp_2}{\lc}{E}}{\Trg_2}{\lc}{V}{\Hp_3}{\Trg_3}
  }{
    \notlocked{\lc}{\SSet}
    &(\lc', \lc) \in \Ord
    &\Hp(\lc) = E
  }
}
\vspace{1ex}\\
\refstepcounter{lab}
\infer[(\thelab\label{evalmore_loc_expr_now})]{   
  \evalexpr{\Hp}{\Trg}{\lc}{V}{\update{\Hp'}{\lc}{V}}{\Trg'}
}{
  \Trg = (\Ord, \SSet)
  &\notlocked{\lc}{\SSet}
  &\notconstrained{\lc}{\Ord}
  &\Hp(\lc) = E
  &\evalexpr{\update{\Hp}{\lc}{\error}}{\Trg}{E}{V}{\Hp'}{\Trg'}
}
\vspace{1ex}\\
\refstepcounter{lab}
\infer[(\thelab\label{evalmore_close})]{   
  \evalexpr{\Hp}{\Trg}{\close{M}}
           {\vbscord{\emp}{o}{\seq{x_i}{i\in I}{\lc''_i}}
                    {\ordclose{\{x_i \mid i \in I\}}{\trg}}}
           {\updateseq{
              \updateseq{
                \updateseq{\Hp_2}{\lc_i}{i\in I}
                          {\substseq{E_i}{x_j}{j\in I}{\lc'_j}}}
                        {\lc'_i}{i\in I}{\lc_i}}
                      {\lc''_i}{i\in I}{\lc_i}}
           {(\addord{\Ord}{\Ord'}, \addord{\SSet}{\SSet'})}
}{
  \deduce{   
    \forall i \in I, ~\lc_i ~\lc'_i ~\lc''_i ~\fresh
    ~~\Ord' = \substseq{\substseq{\substseq{\ord}{x_i}{i\in I}{\lc_i}}
                                  {\ixt{x_i}}{i\in I}{\lc'_i}}
                        {\ext{x_i}}{i\in I}{\lc''_i}
    ~~\SSet' = \substseq{\substseq{\substseq{\sset}{x_i}{i\in I}{\lc_i}}
                                  {\ixt{x_i}}{i\in I}{\lc'_i}}
                        {\ext{x_i}}{i\in I}{\lc''_i}
  }{
    \evalexpr{\Hp}{\Trg}{M}{\vbscord{\emp}{o}{\seq{x_i}{i\in I}{E_i}}{\trg}}
             {\Hp_2}{\Trg_2}
    &\trg = (\ord, \sset)
    &\Trg_2 = (\Ord, \SSet)
  }
}
\end{tabular}}
\caption{Extension with trigger-constraint}\label{eval_disciplined_more_fig}
\end{figure*}

To gain extra expressivity to deal with the trigger-constraint, 
we extend a local constraint to be a pair of a binary relation on identifiers
and a set of sets of identifiers. 
Precisely, a structure \bscord{\iota}{o}{\rho}{\trg} now contains 
a local constraint \trg, which is a pair $(\ord, \sset)$ of a binary relation
\ord\ on $ \dom{\iota} \cup \dom{\rho} \cup \{ \ext{x} \mid x \in \dom{\iota} \cup \dom{\rho} \} \cup \{ \ixt{x} \mid x \in \dom{\iota} \cup \dom{\rho} \}$ and a set \sset\ of sets of elements in $\dom{\iota}\cup \dom{\rho}$. 
$\ord$ expresses
constraints on the evaluation order and accessibilities as before.
\sset\ expresses the trigger-constraint; 
if \mX\ is in \sset, then evaluation of all components 
bound to identifiers in \mX\ is triggered at once when any of the components 
is accessed for the first time. For instance when \sset\ contains
\dom{\rho}, then all defined components of the structure, 
but deferred ones, are evaluated at once when any of
the defined components is accessed for the first time. 
Accordingly, a global constraint \Trg\ is now a pair $(\Ord, \SSet)$
of a binary relation \Ord\ on locations and set \SSet\ of sets of locations.
The functions \iordclose\ and \iordmix\ need to be extended to 
operate on pairs of a binary relation and set of sets.

To take account of the trigger-constraint,
We replace rules (\ref{evalcnstr_loc_expr_later}) 
and (\ref{evalcnstr_loc_expr_now}) by 
rules (\ref{evalmore_loc_expr_trigger}), 
(\ref{evalmore_loc_expr_later}) and (\ref{evalmore_loc_expr_now}),
and rule (\ref{evalcnstr_close}) by rule  (\ref{evalmore_close}) as given
in Figure~\ref{eval_disciplined_more_fig}. 
The notation \notlocked{\lc}{\SSet} denotes the condition that
\SSet\ does not contain a set containing \lc. 

The additional work is to check, before evaluating a location \lc,
whether there are locations whose evaluation is triggered by \lc. 
Rule (\ref{evalmore_loc_expr_trigger}) considers the case where \lc\ 
triggers evaluation of $\lc_i$'s, as described by the side condition 
$\{ \lc, \lc_1 \ldots \lc_n \} \in \SSet$. The rule bears responsibility
for evaluating $\lc_i$'s and  $\{ \lc, \lc_1 \ldots \lc_n \}$ 
is discharged from \SSet.
Except for the side condition \notlocked{\lc}{\SSet},
rules (\ref{evalmore_loc_expr_later}) and (\ref{evalmore_loc_expr_now})
are identical to previous rules
(\ref{evalcnstr_loc_expr_later}) and 
(\ref{evalcnstr_loc_expr_now}); the side condition ensures 
that all trigger constraints involving \lc\ have been handled. 
It is easily proved that 
(\ref{evalcnstr_loc_val})
and (\ref{evalmore_loc_expr_trigger}) are exclusive to each other. 
Rule (\ref{evalmore_close}) is a technical adjustment, 
since the rule now needs to 
transport a pair of a binary relation and set of sets 
from the local constraint to the 
global constraint. 
The notation 
\substseq{\substseq{\substseq{\sset}{x_i}{i\in I}{\lc_i}}
                                    {\ixt{x_i}}{i\in I}{\lc'_i}}
                                    {\ext{x_i}}{i\in I}{\lc''_i}
denotes the set of sets of locations obtained from \sset\ 
by substituting $\lc_i$'s for $x_i$'s , $\lc'_i$'s for \ixt{x_i}'s, and 
$\lc''_i$'s for \ext{x_i}'s.
We do not repeat the other rules; 
the necessary modification is to replace \Ord's by \Trg's.

\subsubsection{An example}
As an example of how to use the trigger-constraint, 
we present a possible object initialization strategy 
in class-based object-oriented languages. 
The purpose of the example is not to explain an encoding of
object-oriented language features such as inheritance and
overriding, which is already examined in previous work~\cite{cms}. 
Here we focus on the aspect of object initialization,
by regarding components of a mixin as (instance) fields of an object. 
In this scenario, open mixins correspond to classes 
and closed mixins to objects. 
We shall not need the full expressivity of \nickname\ to model standard
class-based object-oriented languages. 
For specificity,  we restrict ourselves to a fragment of
the calculus satisfying the following conditions: 
\begin{itemize}\itemsep=0ex
\item Structures do not contain sub-mixins. 
This implies we do not consider inner classes.
\item The sum operation only takes open mixins as operands. 
The operation corresponds to inheritance. Then this is standard, 
since usually classes do not inherit from objects, or vice versa.
In the sum construct \mix{M_1}{M_2}, 
we assume the left operand mixin $M_1$ corresponds to the superclass
and the right operand mixin $M_2$ to the inheriting class. 
\end{itemize}

For a structure \bscord{\iota}{o}{\rho}{(\ord, \sset)}, 
we assign internally-lazy-record
externally-lazy-record strategy. The initialization order of fields
within an object is kept unspecified. 
That is, we let $\ord$ be the relation:\vspace{1ex}\\
\hspace*{1ex}$\begin{array}{l}
\hspace{2ex}\{(x_i, \ext{x_j}) \mid x_i, x_j  \in (\dom{\iota} \cup \dom{\rho}) \} \\
\cup\ \{(x_i, \ixt{x_j}) \mid x_i, x_j  \in (\dom{\iota} \cup \dom{\rho}) \}
\end{array}$\vspace{1ex}\\
We let $\sset$ be the singleton:\vspace{1ex}\\
\hspace*{3ex}$\{ (\dom{\iota} \cup \dom{\rho}) \}$
\vspace{1ex}\\
The trigger-constraint \sset\ above stipulates that all the 
fields of an object must be initialized at once. 


We give interpretations to functions \iordclose\ and \iordmix\ as follows:\vspace{1ex}\\
\hspace*{1ex}$\begin{array}{l}
\ordclose{\mX}{\trg} = (\emp, \emp)\\
\ordmix{\mX}{(\ord, \sset)}{\mX'}{(\ord', \sset')} = \\
~~~~(\ord \cup\ \ord' \cup\ \{(x_i, x_j) \mid x_i \in \mX, x_j \in \mX' \}, 
~\{(\mX \cup \mX')\})
\end{array}$\vspace{1ex}\\

\begin{figure}
{\sf \begin{tabular}{l}
class A \{ val a1 = ...  \} \\
class B extends A \{ val b1 = ... val b2 = ... \}\\
class C extends B \{ val c1 = .. \}\\
let c = new C\\
\end{tabular}}
\caption{A class hierarchy}\label{ex:class_hierarchy}
\end{figure}

The interpretation of \iordmix\ makes the strategy interesting.
Viewed as object initialization, the strategy is described as follows.
\begin{enumerate}\itemsep=0ex
\item Fields are initialized following the class hierarchy. 
That is, fields of superclasses are initialized before those of sub-classes. 
For instance in Figure~\ref{ex:class_hierarchy}, 
the field {\sf a1} of the object {\sf c} is initialized
before {\sf b1} or {\sf b2}, and {\sf b1} and {\sf b2} are before {\sf c1}. 
This is imposed by including the constraint 
$\{(x_i, x_j) \mid x_i \in \mX, x_j \in \mX' \}$ in the result of the first element 
of \iordmix. 

\item Any field inherited from a superclass 
becomes accessible both inside and outside, once all the fields 
from the superclass are initialized. 
Hence in the example, {\sf c.b1} and {\sf c.b2} become 
accessible immediately after both have 
been initialized, but before {\sf c.c1} is initialized. 
Note that neither of {\sf c.b1} nor {\sf c.b2} is accessible even inside
before both have been initialized. 
This is specified by the local constraint assigned to a structure. 

\item We assume the {\sf new} operation for instantiating objects from classes
is equivalent to the close operation 
followed by access to some field  of the mixin closed.
For instance we may assume the {\it Object} class which resides 
in the root of the class hierarchy, i.e. a superclass of any class,
contains a special field named {\sf init} for that purpose. 
Then the strategy enforces the standard object initialization policy
where all fields of an object are initialized when it is created. 
This is where we use the trigger-constraint.
The result $\{(\mX \cup \mX')\}$ of the second element of 
\iordmix\ ensures that 
all field of an object are initialized at once when {\sf init} is accessed,
since it triggers initialization of the fields of inheriting classes, when
the {\sf init} field is initialized. 
\end{enumerate}

In fact this strategy, in particular the characteristic described second above,
is inspired by the object initialization strategy
of the \FSharp\ programming language~\cite{FSharp}~\footnote{We believe 
the strategy presented is close to \FSharp's. Yet the exact strategy is 
under-specified in the language documentation.}.
\FSharp\ places restrictions on possible object initialization 
patterns, enhancing safety by eliminating 
programming styles which are often awkward sources
for null-pointer exceptions. 
As a result, the strategy is less flexible than that of some object-oriented
languages such as Java, yet it can still
support common programming idioms found in object-oriented programming. 
In short, we found \FSharp's strategy an interesting example of disciplined 
evaluation orders.

\section{Related Work}\label{related_sec}

One difference between previous work and the present work
is that each of previous work examined an evaluation strategy,
while we explored the design space of lazy evaluation strategies 
and give the operational semantics which can deal with several strategies. 
Most related to our work is Ancona and Zucca's call-by-name mixin
calculus~\cite{cms} and Hirschowitz and Leroy's call-by-value mixin
calculus~\cite{cbvmixins}.  
Ancona et al. investigated the interaction between mixins
and computational effect, using a monadic metalanguage as semantic 
basis~\cite{cmsdo}. We will review the three in detail below. 
S. Fagorzi and E. Zucca proposed $R$-calculus to allow projection 
from open mixins in a consistent way\\ 
\cite{openmodules}; 
this is a design direction we have not explored. 
Our previous work proposed a lazy evaluation strategy for 
recursive ML-style modules, which we explained in Section~\ref{sec:extension}. 
Design questions we encountered there motivated 
strategies we proposed in Section~\ref{sec:strategies}. 
We do not look back at the history of mixins, but only mention 
a few of influential papers~\cite{jigsaw,cms,cbvmixins,mixinmodules,cool&hot}. 
In particular, type systems which eliminate unsound scenarios
such as to close mixins having holes or to select a component 
from an open mixin, have been well-investigated. 
Those type system proposals are orthogonal and complementary to 
the presented work. 

\vspace{.5ex}

{\it Call-by-name mixins}
Ancona and Zucca formalized a mixin calculus, named \CMS,  
with call-by-name evaluation using small-step semantics \cite{cms}.
The formalization is concise and allows equational reasoning of mixins. 
However the call-by-name semantics might not 
be suitable to be used with a call-by-value 
core language allowing arbitrary side-effects, 
such as the ML core language, since 
with the call-by-name semantics evaluation of {\sf let rec x = x}
diverges when {\sf x} is selected and 
the same side-effect can be produced repeatedly. 
Large part of the formalization in Section~\ref{sec:lazymixins} is 
borrowed from \CMS. We adapted their small-step semantics 
to big-step semantics. 
Technically we made a small but important modification;
in \CMS\ an output assignment maps names to expressions, whereas
in \nickname\ names to identifiers. 
In this way we avoid duplicating expressions in the freeze operation.

\vspace{.5ex}

{\it Call-by-value mixins}
Hirschowitz and Leroy formalized a mixin calculus 
with call-by-value evaluation~\cite{cbvmixins}.
Generally call-by-value mixins result in the simplest evaluation order, 
while lazy mixins are more flexible in handling recursion. 
For instance, the marshallers example in Section~\ref{sec:picklers}
relies on the laziness to specify marshallers for recursive data types. 
We are motivated to distinguish open and closed mixins 
by their work. 
Apart from the difference of call-by-value and lazy,
\nickname\ differs from their calculus in the sum operation 
in that we allow mixins to be merged independently of whether they are
closed or open, 
while they only consider the sum operation on open mixins. 
Our design choice is motivated to keep flexibility in sharing side effects. 
For instance, it was useful in {\sf MakeSet}
and {\sf MakeMultiSet} mixins example from Section~\ref{sec:makeset};
the sharing of {\sf counter} between {\sf Set} and {\sf MultiSet} is achieved
by merging the closed mixin {\sf Key} with {\sf MakeSet} 
and {\sf MakeMultiSet}. 
It should be noted that one important objective of Herschowitz and Leroy's work is to 
statically ensure initialization safety of mixins.
The objective of our paper is not about static guarantees of initialization safety.

\vspace{.5ex}

{\it Effectful mixins and equational reasoning}
Ancona et al. examined the interaction between mixins and 
computational effect, by means of a recursive monadic binding~\cite{cmsdo}.
The sum operation only takes open mixins as argument in their calculus.
They separate computational components from non-computational ones,
where the former are evaluated exactly once, while the latter are re-evaluated 
whenever they are projected. Computational components of a mixin 
are evaluated at once  when the mixin is closed by the {\it doall} operation, 
in a similar way to the rule (\ref{evalcbv_close}) from Section~\ref{variants_sec}. 
To the best of our understanding, their evaluation strategy is top-down
internally-lazy-field externally-lazy-record strategy, 
where closed sub-mixins are evaluated immediately. When mixins are merged,
the computational components of the right-operand mixin are 
evaluated before those of the left-operand mixin. 
The separation of computational components lets them retain the 
\CMS\ equational reasoning.

\section{Conclusion}\label{conclude_sec}

We have formalized the operational semantics 
for a mixin calculus with lazy evaluation. 
We started by considering a simpler semantics
where a component of a mixin is evaluated when it is projected. 
Then we extended the semantics to impose various restrictions
on the evaluation order of and accessibilities to components of mixins, 
by adding local constraint to mixin structures
and global constraint to the evaluation judgment. 
We have kept the formalization neutral of constraints
and it can be instantiated to express several evaluation strategies. 
As well as we considered the design space of strategies, 
we took a closer look at two particular strategies. 

We do not claim the formalization is expressive enough to deal with
all interesting evaluation strategies. 
At the same time, 
we do not think it technically difficult to extend the formalization 
to express more strategies.
As the extension we made in Section~\ref{sec:extension} exemplifies,
we may well gain more expressivity by extending the constraint language and
by adding more evaluation rules.
Indeed a moot point was to keep the formalization not too complicated 
without giving up too much expressivity to deal with interesting strategies. 


\paragraph{Acknowledgement}
I am grateful to Xavier Leroy for the valuable advice and fruitful discussions
throughout the development of this work.

\bibliographystyle{plainnat}
\bibliography{references}

\end{document}